\documentclass[twoside,12pt]{article}
\usepackage{extsizes}
\usepackage[super,sort&compress,comma]{natbib} 
\usepackage[version=3]{mhchem}
\usepackage[left=1.5cm, right=1.5cm, top=1.785cm, bottom=2.0cm]{geometry}
\usepackage{balance}
\usepackage{mathptmx}
\usepackage{amsmath,amsfonts,amssymb,amsthm} %
\usepackage{sectsty}
\usepackage{graphicx} 
\usepackage{epstopdf}
\usepackage{lastpage}
\usepackage[format=plain,justification=justified,singlelinecheck=false,font={stretch=1.125,small,sf},labelfont=bf,labelsep=space]{caption}
\usepackage{float}
\usepackage{fancyhdr}
\usepackage{fnpos}
\usepackage[english]{babel}
\addto{\captionsenglish}{%
  \renewcommand{\refname}{Notes and references}
}
\usepackage{array}
\usepackage{droidsans}
\usepackage{charter}
\usepackage[T1]{fontenc}
\usepackage[usenames,dvipsnames]{xcolor}
\usepackage{setspace}
\usepackage[compact]{titlesec}
\usepackage{hyperref}

\begin{document}

\title{Range separation of the Coulomb hole$^{\dag\ddag}$}

\renewcommand{\thefootnote}{\fnsymbol{footnote}}

\author{Mireia Via-Nadal,\textit{$^{a,b}$} Mauricio Rodr\'iguez-Mayorga,\textit{$^{a,c}$} Eloy Ramos-Cordoba,\textit{$^{a,b}$}$^\ast$ \\  and Eduard                Matito\textit{$^{a,d}$}$^\ast$}

\maketitle

\begin{abstract}
	A range-separation of the Coulomb hole into two components, one of them being predominant at long interelectronic separations ($h_{\text{c}_{I}}$) and the other at short distances ($h_{\text{c}_{II}}$), is exhaustively analyzed throughout various examples that put forward the most relevant features of this approach and how they can be used to develop efficient ways to capture electron correlation. We show that $h_{\text{c}_{I}}$, which only depends on the first-order reduced density matrix, can be used to identify molecules with a predominant nondynamic correlation regime and differentiate between two types of nondynamic correlation, types A and B.
	Through the asymptotic properties of the hole components, we explain how $h_{\text{c}_{I}}$ can retrieve the long-range part of electron correlation. We perform an exhaustive analysis of the hydrogen molecule in a minimal basis set, dissecting the hole contributions into spin components. We also analyze the simplest molecule presenting a dispersion interaction and how $h_{\text{c}_{II}}$ helps identify it. The study of several atoms in different spin states reveals that the Coulomb hole components distinguish correlation regimes that are not apparent from the entire hole. The results of this work hold the promise to aid in developing new electronic structure methods that efficiently capture electron correlation.
\end{abstract}

\renewcommand*\rmdefault{bch}\normalfont\upshape
\rmfamily
\section*{}
\vspace{-1cm}

\footnotetext{\textit{$^{a}$~Donostia International Physics Center (DIPC), 20018 Donostia, Euskadi, Spain.}}
\footnotetext{\textit{$^{b}$~Kimika Fakultatea, Euskal Herriko Unibertsitatea (UPV/EHU), Donostia, Euskadi, Spain.}}
\footnotetext{\textit{$^{c}$~
Department of Theoretical Chemistry, VU University Amsterdam, De Boelelaan 1083, 1081 HV Amsterdam, The Netherlands}}
\footnotetext{\textit{$^{d}$~IKERBASQUE, Basque Foundation for Science, 48013 Bilbao, Euskadi, Spain.}}
\footnotetext{\ddag~This paper is dedicated to Paul Geerlings on occasion of his 70th birthday.}
\footnotetext{\dag~Electronic Supplementary Information (ESI) available: The Supporting Information contains the Coulomb hole of the He-Ne atomic series, a study of the basis set consistency for the Be atom, the short-range region of the Coulomb hole of the beryllium molecule, a spin decomposition of the Coulomb hole of H$_2$ in a minimal basis set, and a more detailed development of the asymptotics and conditions of $\Delta\rho_2^{\text{c}_{\textit{I}}} (\boldsymbol{1},\boldsymbol{2})$ and $\Delta\rho_2^{\text{c}_{\textit{II}}} (\boldsymbol{1},\boldsymbol{2})$ for the hydrogen molecule in a minimal basis. See DOI: 10.1039/cXCP00000x/}

\newcommand*\mycommand[1]{\texttt{\emph{#1}}}
\newcommand{\cc}[1]{{\color{red}#1\color{black}}}
\newcommand{\HH}{{\text{H}}}

\section{Introduction}
In computational chemistry, the difference between the exact nonrelativistic electronic energy and the restricted Hartree--Fock (RHF) one is known as correlation energy.\cite{lowdin:55pr} Even though the uncorrelated, self-consistent field (SCF) HF energy usually represents more than 98\% of the total energy, the remaining is crucial to describe the chemistry of atoms and molecules (dissociation energies, reaction enthalpies, etcetera). \cite{szabo:12book,wilson:07book} The description of a quantum system with the HF method lacks \textit{electron correlation}, \textit{i.e.}, a correct account of the correlated motion of the electrons. This conundrum is generally known as the many-body problem, and it is one of the main challenges in this field.\cite{szabo:12book,wilson:07book} 
The study of electron correlation \textit{per se} goes hand in hand with developing efficient electronic structure methods.\cite{hattig:12cr,tew:07jcc,cremer:01mp,ziesche:00the,raeber:15pra,benavides:17pra,gottlieb:05prl,hollett:11jcp,fogueri:13tca,proynov:13pra,cioslowski:91pra,valderrama:97jcp,valderrama:99jcp,mok:96jpc,benavides:17pccp,vuckovic:17pccp,ramos-cordoba:16pccp,ramos-cordoba:17jctc,via-nadal:19jpcl}\newline

Many types of electron correlation exist, nondynamic (or static)~\cite{bartlett:07rmp,sinanoglu:64acp,szabo:12book} and dynamic~\cite{sinanoglu:64acp,szabo:12book} being the most regularly used because electronic structure methods are usually classified according to their ability to retrieve them. Dynamic correlation is universally present in systems with at least two electrons, as it describes the motion of charged particles avoiding each other due to the electronic repulsion. Hence, this type of correlation increases with the number of electrons. A reference single-determinant picture along with a large number of low-contributing configurations is usually sufficient to portray this contribution. It is thus natural that the electron density displays very small differences with respect to the reference HF density.\cite{cioslowski:91pra,valderrama:97jcp,via-nadal:19jpcl} Due to its nature, dynamic correlation affects electrons that are close to each other (short-ranged), but it is also responsible for long-range dispersion forces.~\cite{stone:13book}
Configuration interactions or coupled cluster with single and double excitations (CISD or CCSD), \cite{shavitt:77ci,cizek:66jpc} Møller-Plesset second-order perturbation theory (MP2)~\cite{moller:34pr} and density functional approximations (DFAs)~\cite{kohn:65pr,cremer:01mp} are methods that retrieve a large amount of dynamic correlation. On the other hand, nondynamic correlation is not universal and emerges in the presence of partially-occupied (near-)degenerate orbitals. It is characteristic of bond stretching, polyradical structures, entanglement, and high symmetries. The correct description of nondynamic correlation requires a wavefunction that mixes other largely-contributing configurations besides the HF ground-state one.~\cite{helgaker:00book} Nondynamic correlation induces considerable changes in the electron density caused by the mix of highly-contributing configurations in the CI vector. \cite{cioslowski:91pra,valderrama:97jcp,valderrama:99jcp,ramos-cordoba:16pccp,ramos-cordoba:17jctc,via-nadal:19jpcl} Complete active space SCF (CASSCF),~\cite{roos:87acp} density matrix renormalization group (DMRG),~\cite{white:93prb} and multi-configurational SCF (MCSCF) \cite{frenkel:34mcscf,roos:77mcscf} are methods that can retrieve a fair amount of nondynamic correlation. As we shall confirm later, nondynamic correlation is important both at short and long range.\newline

Systems presenting dynamic and nondynamic correlation have become one of the greatest challenges in modern electronic structure methods. The ability to simultaneously tackle both types of correlation is present in very few electronic structure methods, such as the complete active space with second-order perturbation theory (CASPT2),~\cite{andersson:90jpc,andersson:92jcp} multireference configuration interactions (MRCI)~\cite{buenker:74tca,werner:88jcp} or, most recently, the adiabatic-connection MCSCF (AC-MCSCF)~\cite{pastorczak:19jpcl}, $\Delta$NO,~\cite{hollett:20jcp} and GNOF.~\cite{piris:21prl} Nevertheless, these methods are far from exact and their computational cost represents a big drawback. Latterly, an interest in hybrid schemes such as range-separated methods~\cite{savin:96bc,savin:88ijqc} has surged to confront the exposed problem. These methods aim to recover both correlation types by treating short- and long-range interactions with two different methodologies, according to their ability to recover one of the correlation components.~\cite{savin:88ijqc, fromager:07jcp, toulouse:09prl, toulouse:04pra, iikura:01jcp, baer:10arpc, garrett:14jctc, savin:96bc,bao:17pccp, gagliardi:17acr, li:14jctc} Although range-separation hybrid schemes provide a splitting of the pair density and the interelectronic coordinate, the separation is not custom-built for the correlation type present in the system. \newline

Some studies have put forward measures for quantifying dynamic and nondynamic correlation,~\cite{raeber:15pra,benavides:17pra,gottlieb:05prl,hollett:11jcp,fogueri:13tca,proynov:13pra,cioslowski:91pra,valderrama:97jcp,valderrama:99jcp,mok:96jpc,benavides:17pccp,vuckovic:17pccp} some of which are based on the correlation energy.~\cite{cioslowski:91pra,valderrama:97jcp,valderrama:99jcp,mok:96jpc,benavides:17pccp,vuckovic:17pccp} As reported by Coleman, \cite{coleman:67ijqc} the use of electron-pair distribution functions to study electron correlation must lead to an understanding of both short- and long-range interelectronic interactions. In line with the former statement, we have recently proposed a general decomposition scheme to separate the correlation part of the pair density into two components that permit the identification of systems with prevalent dynamic or nondynamic correlation.~\cite{ramos-cordoba:16pccp,ramos-cordoba:17jctc,via-nadal:19jpcl} 
From this scheme, scalar~\cite{ramos-cordoba:16pccp} and local~\cite{ramos-cordoba:17jctc} measures of dynamic and nondynamic correlation have been developed. Finally, a range-separation of the Coulomb hole has been introduced, providing the dominance of one component at long ranges (c$_{I}$) and the other one (c$_{II}$) at short ranges. c$_I$ ---which only requires the first-order reduced density matrix (1-RDM)--- has been shown to reflect nondynamic correlation, whereas c$_{II}$ includes both correlation types.~\cite{via-nadal:19jpcl}\newline

In the present work, the range separation of the Coulomb hole is exhaustively analyzed throughout various examples that put forward the most relevant features of this approach and how they can be used to develop efficient ways to capture electron correlation.
First of all, we introduce the range-separation of the Coulomb hole and its rationalization. Through the asymptotic properties of the hole parts, we explain how the component based on the first-order reduced density matrix can retrieve the long-range part of the correlated pair density. Second, we perform an exhaustive analysis of the hydrogen molecule in a minimal basis set, dissecting the hole contributions into spin components. Third, we analyze the simplest molecule presenting a dispersion interaction and how one of the Coulomb hole components helps identify it. This dispersion signature is also present in the remaining molecules of the manuscript, highlighting its universal character. Afterward, we analyze the Coulomb holes of several atoms in different spin states, finding that the Coulomb hole components distinguish correlation regimes that are not apparent from the entire hole. Finally, we analyze the two types of nondynamic correlation, types A and B,\cite{hollett:11jcp} and show that the Coulomb hole's first component can capture them.\newline

The results of this work hold the promise to aid in developing new electronic structure methods that efficiently capture electron correlation. In particular, the models of the $c_{II}$ component can be straightforwardly used in the reduced density matrix functional theory,\cite{piris:14ijqc,pernal:15tcc,ramos-cordoba:15jcp,ramos-cordoba:14jcp,cioslowski:15jcp} although they are not limited to this theory.

\section{Methodology}

The pair density of an $N$-electron system described by the $\Psi(\boldsymbol{1},\dots,\boldsymbol{N})$ wavefunction is

\begin{equation}
\rho_2 (\boldsymbol{1},\boldsymbol{2}) =  N(N-1) \int \text{d}_{\boldsymbol{3}}\dots \text{d}_{\boldsymbol{N}} |\Psi(\boldsymbol{1},\dots,\boldsymbol{N})|^2,
\label{eqn:spinlesspairdensity}
\end{equation}

where we have assumed McWeeny's normalization, \cite{mcweeny:89book} which accounts for the number of  electron pairs in the system, and the variables $\left(\boldsymbol{1},\boldsymbol{2},\ldots \right)$ refer to both space and spin coordinates, $\boldsymbol{1}\equiv\vec{r}_1, \sigma_1 $. The pair density contains information about the probability of finding electron $1$ at the position and spin $\boldsymbol{1}$ and electron $2$ at the position and spin $\boldsymbol{2}$ when the position of the remaining of $N-2$ electrons is averaged over the whole space. The analysis of the pair density is arduous since it depends on six space and two spin variables. Conversely, the radial intracule density is a reduced form of the pair density that only depends on the interelectronic distance or range separation, $s$, and still retains the necessary information to calculate the electronic repulsion energy,

\begin{equation}
V_{ee} = \frac{1}{2} \int \text{d}s \frac{I(\rho_2,s)}{s},
\label{eqn:vee}
\end{equation}

being

\begin{equation}
I(\rho_2,s) = \int \text{d}_{\boldsymbol{1}} \text{d}_{\boldsymbol{2}}\, \rho_2 (\boldsymbol{1},\boldsymbol{2}) \delta(s-r_{12}),
\label{eqn:intracule}
\end{equation}
where $r_{12}$ is the Euclidean distance between electrons 1 and 2, and $\delta(s-r_{12})$ is the Dirac delta. Eq~\ref{eqn:intracule} is the radial intracule density and it corresponds to the probability distribution of interelectronic distances $s$ between two electrons. It provides information about the electron-pair relative motion within atoms and molecules, being also a valid quantity to interpret the electronic structure of atoms and chemical  bonding.\cite{rodriguez:18jpca,sarasola:90jpb,sarasola:92jcp,mercero:17cjc,rodriguez:19co}
This distribution can also be obtained experimentally from X-ray scattering cross-sections.~\cite{thakkar:78jpb, thakkar:84ijqc, thakkar:84pra, valderrama:01pra}\newline

After Löwdin's definition of correlation energy,~\cite{lowdin:55pr} Coulson defined the correlated pair density as the difference between the exact and the HF pair densities,~\cite{coulson:60rmp}
\begin{equation}
\Delta\rho_2^{\text{c}} (\boldsymbol{1},\boldsymbol{2}) = \rho_2(\boldsymbol{1},\boldsymbol{2}) - \rho_2^{\text{HF}}(\boldsymbol{1},\boldsymbol{2}).
\label{eqn:densitycorr}
\end{equation}

The correlated pair density intracule $\Delta\rho_2^{\text{c}} (\boldsymbol{1},\boldsymbol{2})$ is known as Coulson's Coulomb hole: \cite{coulson:61ppsl,boyd:73jpb,valderrama:00book}
\begin{equation}
h_{\text{c}} \left(s\right) = 
I\left(\Delta\rho_2^{\text{c}},s\right)=
I\left(\rho_2,s\right) - I\left(\rho_2^{\text{HF}},s\right).
\label{eqn:coulombhole}
\end{equation}

The Coulomb hole is thus a probability density difference that reflects the effect of switching from the mean-field HF approximation, which underestimates the electronic repulsion, to a correlated framework. As a result, the average interelectronic distance increases from HF to the exact description, $\langle s \rangle_{\text{HF}} < \langle s \rangle$ ($\langle s \rangle = \int I(s) s ds$); this effect is reflected by the negative values of the Coulomb hole at small values of $s$ and positives values at large $s$.\newline

We have introduced elsewhere \cite{ramos-cordoba:16pccp,ramos-cordoba:17jctc,via-nadal:19jpcl} a splitting of the correlated pair density, eq~\ref{eqn:densitycorr}, using the single-determinant approximation to the pair density, \cite{lowdin:55pr} 
\begin{equation}
\rho_2^{\text{SD}}\left(\rho_1,\boldsymbol{1},\boldsymbol{2}\right) = \rho(\boldsymbol{1})\rho(\boldsymbol{2})-\vert\rho_1\left(\boldsymbol{1};\boldsymbol{2}\right)\vert^2,
\label{eqn:SDansatz}
\end{equation}
where $\rho_1\left(\boldsymbol{1};\boldsymbol{2}\right)$ is the first-order reduced density matrix (1-RDM),
\begin{equation}
\rho_1 \left( \boldsymbol{1};\boldsymbol{1'} \right) = \int \text{d}_{\boldsymbol{2}}\ldots\text{d}_{\boldsymbol{N}} \Psi\left(\boldsymbol{1},\boldsymbol{2},\ldots,\boldsymbol{N}\right) \Psi^*\left(\boldsymbol{1'},\boldsymbol{2},\ldots,\boldsymbol{N}\right),
\label{eqn:dm1}
\end{equation}
and its diagonal part, $\rho\left(\boldsymbol{1}\right)=\rho\left(\boldsymbol{1},\boldsymbol{1}\right)$, is the electron density. The SD ansatz, eq~\ref{eqn:SDansatz}, takes advantage of the HF expression for the pair density but uses the actual 1-RDM to generate an approximation to the pair density. Obviously, it returns the HF pair density when the HF 1-RDM is used, \textit{i.e.},

\begin{equation}
\rho_2^{\text{HF}} \left(\boldsymbol{1},\boldsymbol{2}\right) = \rho_2^{\text{SD}}\left(\rho_1^{\text{HF}},\boldsymbol{1},\boldsymbol{2}\right).
\label{eqn:HFdm2}
\end{equation}
The SD approximation includes neither short- \cite{rodriguez:17pccp2} nor long-range \cite{via-nadal:17pra} dynamic correlation, that is, it only considers nondynamic correlation. Even though it cannot be guaranteed that \textit{all} nondynamic correlation in the system is accounted for, it is legitimate to claim that the SD pair density, eq~\ref{eqn:SDansatz}, includes some extent of it. In fact, the SD approximation captures most long-range electron correlation effects and usually presents a minimal short-range contribution that arises from the relative degeneracy of its frontier orbitals (except for systems with type B nondynamic correlation,\cite{hollett:11jcp} \textit{vide infra}). 
 Unlike the HF pair density, the SD approximation is usually not $N$-representable by construction, \textit{i.e.}, this approximate pair density does not necessarily correspond to an $N$-particle fermionic wavefunction. The violation of the $N$-representability conditions may lead to spurious energies \cite{coleman:00book} and affect density matrix properties as the trace, $\text{Tr}\left[\rho_2^{\text{SD}}\right]$, or the positivity of its eigenvalues, among others.~\cite{coleman:63rmp,mazziotti:12prl,rodriguez:17pccp2,rodriguez:17pccp,ayers:07acp,feixas:14jctc} Although it is not $N$-representable, $\rho_2^{\text{SD}} \left(\boldsymbol{1},\boldsymbol{2}\right)$ is used in this context as a mathematical approximation to seize long-range correlation. Indeed, we have used it to separate eq~\ref{eqn:densitycorr} into two correlation contributions, $\Delta \rho_2^{\text{c}_{\textit{I}}}$ and $\Delta \rho_2^{\text{c}_{\textit{II}}}$,\cite{via-nadal:19jpcl}
\begin{eqnarray}
\Delta\rho_2^{\text{c}_{\textit{I}}} (\boldsymbol{1},\boldsymbol{2}) &=& \rho_2^{\text{SD}}\left(\rho_1,\boldsymbol{1},\boldsymbol{2}\right) - \rho_2^{\text{SD}}\left(\rho_1^{\text{HF}},\boldsymbol{1},\boldsymbol{2}\right)
\label{eqn:corr_type_CI}\\
\Delta\rho_2^{\text{c}_{\textit{II}}} (\boldsymbol{1},\boldsymbol{2}) &=& \rho_2\left(\rho_1,\boldsymbol{1},\boldsymbol{2}\right) - \rho_2^{\text{SD}}\left(\rho_1,\boldsymbol{1},\boldsymbol{2}\right),
\label{eqn:corr_type_CII}
\end{eqnarray}
which, along with the HF pair density, recover the exact pair density:
\begin{equation}
\rho_2 \left(\boldsymbol{1},\boldsymbol{2}\right) = 
\rho_2^{\text{HF}}\left(\boldsymbol{1},\boldsymbol{2}\right) + 
\Delta\rho_2^{\text{c}_{\textit{I}}}\left(\boldsymbol{1},\boldsymbol{2}\right) +
\Delta\rho_2^{\text{c}_{\textit{II}}}\left(\boldsymbol{1},\boldsymbol{2}\right).
\end{equation}

The partition of the pair density into the HF reference plus the $c_I$ component and the $c_{II}$ part is known as the Lieb partitioning of the pair density.\cite{lieb:81prl,levy:87bc,buijse:91thesis}
$\Delta\rho_2^{\text{c}_{\textit{II}}} \left(\boldsymbol{1},\boldsymbol{2}\right)$ is also known as the cumulant of the density matrix and captures the correlation lacking in the 2-RDM.\cite{kutzelnigg:99jcp,mazziotti:98cpl} $\Delta\rho_2^{\text{c}_{\textit{I}}} \left(\boldsymbol{1},\boldsymbol{2}\right)$ only depends on the actual 1-RDM and the HF one measures the dissimilarity between these two matrices. The electron density of systems with nondynamic correlation presents more considerable differences with respect to HF than the density of systems with dynamic correlation.\cite{valderrama:97jcp} Therefore, $\Delta\rho_2^{\text{c}_{\textit{I}}} \left(\boldsymbol{1},\boldsymbol{2}\right)$ can be regarded as a function measuring nondynamic correlation, whereas the information about dynamic correlation is expected to be fully contained in $\Delta\rho_2^{\text{c}_{\textit{II}}} \left(\boldsymbol{1},\boldsymbol{2}\right)$, which also includes nondynamic correlation to some extent.
\newline

\begin{figure}
  \centering
  \includegraphics[width=1.0\linewidth]{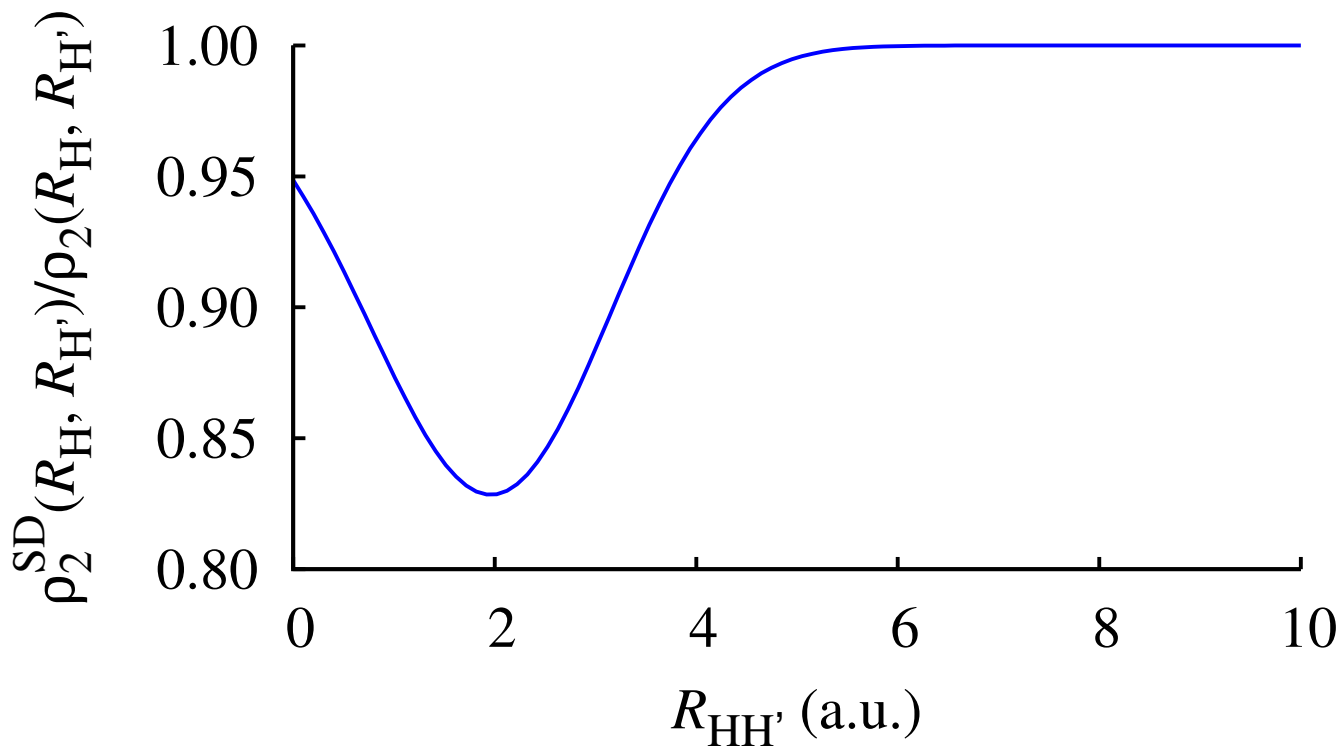} \caption{$\rho_2^{\text{SD}}(R_{\HH},R_{\HH'})/\rho_2(R_{\HH},R_{\HH'})$ against the interatomic separation $R_{\HH\HH'}$ for the minimal basis set calculation of the hydrogen molecule.}
  \label{fig:ratio}
\end{figure}

A more explicit expression for eqs~\ref{eqn:corr_type_CI} and \ref{eqn:corr_type_CII} can be cast:
\begin{eqnarray}
& \hspace{-0.6cm} \Delta\rho_2^{\text{c}_{\textit{I}}} \left(\boldsymbol{1},\boldsymbol{2}\right) =
\vert\rho_{1}^{\text{HF}}\left(\boldsymbol{1};\boldsymbol{2}\right)\vert^2 - 
\vert\rho_{1}\left(\boldsymbol{1};\boldsymbol{2}\right)\vert^2  
+ \rho (\boldsymbol{1}) \rho (\boldsymbol{2}) -  
\rho^{\text{HF}} (\boldsymbol{1}) \rho^{\text{HF}} (\boldsymbol{2}) \hspace{0.6cm}
\label{eqn:corr_type_CI_2}\\
& \Delta\rho_2^{\text{c}_{\textit{II}}} \left(\boldsymbol{1},\boldsymbol{2}\right) =
\rho_2 \left(\boldsymbol{1},\boldsymbol{2}\right) - \rho (\boldsymbol{1})\rho (\boldsymbol{2}) 
+ \vert\rho_{1}\left(\boldsymbol{1};\boldsymbol{2}\right)\vert^2. 
\label{eqn:corr_type_CII_2}
\end{eqnarray}
If we employ the correlation components of the pair density, we can split the Coulomb hole into two components,
\begin{equation}
h_{\text{c}} \left(s\right) = h_{\text{c}_{\textit{I}}} \left(s\right) + h_{\text{c}_{\textit{II}}} \left(s\right)  =
I\left(\Delta\rho_2^{\text{c}_{\textit{I}}},s\right) + I\left(\Delta\rho_2^{\text{c}_{\textit{II}}},s\right),
\end{equation}
which permit the analysis of electron correlation in terms of interelectronic ranges. \cite{via-nadal:19jpcl}
The asymptotic properties of $\Delta\rho_2^{\text{c}_{\textit{I}}} \left(\boldsymbol{1},\boldsymbol{2}\right)$ and $\Delta\rho_2^{\text{c}_{\textit{II}}} \left(\boldsymbol{1},\boldsymbol{2}\right)$ determine the long-range behavior of the Coulomb hole components.
The first important property of $h_{\text{c}_{\textit{I}}} \left(s\right)$ and $h_{\text{c}_{\textit{II}}}\left(s\right)$ is that they vanish for very large values of $s$ as long as 
$s$ is larger than the effective maximum length of the molecule.
Since both the HF and the exact pair density are zero for such points, one only needs
to prove the same for the SD approximation. Two terms form the latter, the first
one involving the product of two densities and, therefore, it vanishes in regions far from
the molecule. The 1-RDM long-range asymptotics~\cite{march:81jcp} also guarantees that the second term,
including the square of a 1-RDM, vanishes under this condition.
Second, we study the behavior of the two Coulomb hole components at short ranges. By the Pauli
principle, the same-spin component of $\Delta\rho_2^{\text{c}_{\textit{I}}} \left(\boldsymbol{1},\boldsymbol{2}\right)$ and $\Delta\rho_2^{\text{c}_{\textit{II}}} \left(\boldsymbol{1},\boldsymbol{2}\right)$ vanish when $r_{12}=0$. Hence, one can easily prove
that $\Delta\rho_2^{\text{c}_{\textit{I}}} \left(\vec{r}_1,\vec{r}_1\right)=2\left(\rho^{\alpha}(\vec{r}_1)\rho^{\beta}(\vec{r}_1)-\rho^{\text{HF},\alpha}(\vec{r}_1)\rho^{\text{HF},\beta}(\vec{r}_1)\right)$. Since HF underestimates the electron-nucleus cusp, this quantity is greater than zero in the vicinity of the nuclei. Our experience indicates that $h_{\text{c}_I}\left(s\right)$ is usually greater than zero everywhere.\footnote{A remarkable exception to this rule occurs in molecules that HF dissociates into fragments with the wrong number of electrons. See the last two examples given in this paper.} Conversely, $h_{\text{c}_{II}}\left(s\right)$  is usually negative at short ranges because $\Delta\rho_2^{\text{c}_{\textit{II}}} \left(\vec{r}_1,\vec{r}_1\right)=2\left(\rho_2^{\alpha\beta}\left(\vec{r}_1,\vec{r}_1\right)-\rho^{\alpha}(r_1)\rho^{\beta}(r_1)\right)$ is negative at points close to the nuclei, which contribute the most to $\Delta\rho_2^{\text{c}_{\textit{II}}} \left(\vec{r}_1,\vec{r}_1\right)$. In the following, we analyze  $\Delta\rho_2^{\text{c}_{\textit{II}}} \left(\boldsymbol{1},\boldsymbol{2}\right)$ at long ranges to prove its short-ranged character. The largest long-range contributions come from two points close to nuclei, typically points that are centered into two different atoms. For the sake of simplicity, let us choose the hydrogen molecule dissociation to illustrate this point. We can take the leading term in the expansion of $\Delta\rho_2^{\text{c}_{\textit{II}}} \left(\boldsymbol{1},\boldsymbol{2}\right)$ around the two electron-nucleus cusps, $\Delta\rho_2^{\text{c}_{\textit{II}}} \left(R_{\HH},R_{\HH'}\right)=\rho_2(R_{\HH},R_{\HH'})-\rho_2^{\text{SD}}(R_{\HH},R_{\HH'})$, which we have fully developed in the Supporting Information. In Figure~\ref{fig:ratio}, we represent the ratio $\rho_2^{\text{SD}}(R_{\HH},R_{\HH'})/\rho_2(R_{\HH},R_{\HH'})$ against the interatomic separation $R_{\HH\HH'}$. The ratio is always greater than 0.8 and easily achieves 1.0 as the bond stretches, indicating that $\Delta\rho_2^{\text{c}_{\textit{II}}} \left(R_{\HH},R_{\HH'}\right)$ quickly vanishes at long distances and, therefore, $\Delta\rho_2^{\text{c}_{\textit{II}}} \left(\boldsymbol{1},\boldsymbol{2}\right)$ is expected to vanish quickly with $r_{12}$ and be predominantly short-ranged.\newline

 Finally, let us mention that the long-range part of $\text{c}_{\textit{II}}$, although very small, it is not exactly zero ---as we shall see in the examples covered in the Results section. $\text{c}_{\textit{II}}$ evaluated at $R$, $R$ being the distance between any two atoms in the molecule will decay like $R^{-3}$ when $R \rightarrow \infty$.\cite{via-nadal:17pra} This dependency is connected with the well-known $R^{-6}$ decay of the van der Waals (vdW) dispersion energy. Because dispersion interactions are weak, the region in $h_{\text{c}_{\textit{II}}}(s)$ is proportionally small.\newline

\section{Computational Details}
Full configuration interactions (FCI) calculations have been run with a modified version of Knowles and Handy's program. \cite{knowles:89cpc,matito:10pccp} Dunning's augmented correlation-consistent double zeta basis set (aug-cc-pVDZ) \cite{kendall:92jcp,dunning:89jcp} is used unless otherwise specified. We have used the basis sets developed in a previous paper~\cite{rodriguez:19co} for the Be isoelectronic series.
The density matrices and the intracule probability distributions have been obtained with the in-house DMN \cite{dmn} and RHO2$_{-}$OPS \cite{rho2ops} codes, respectively; the latter uses the algorithm of Cioslowski and Liu.~\cite{cioslowski:96jcpfast}

\section{Results and discussion}
\subsection{H$_2$ in minimal basis}

\begin{figure}
  \centering
  \includegraphics[width=1\linewidth]{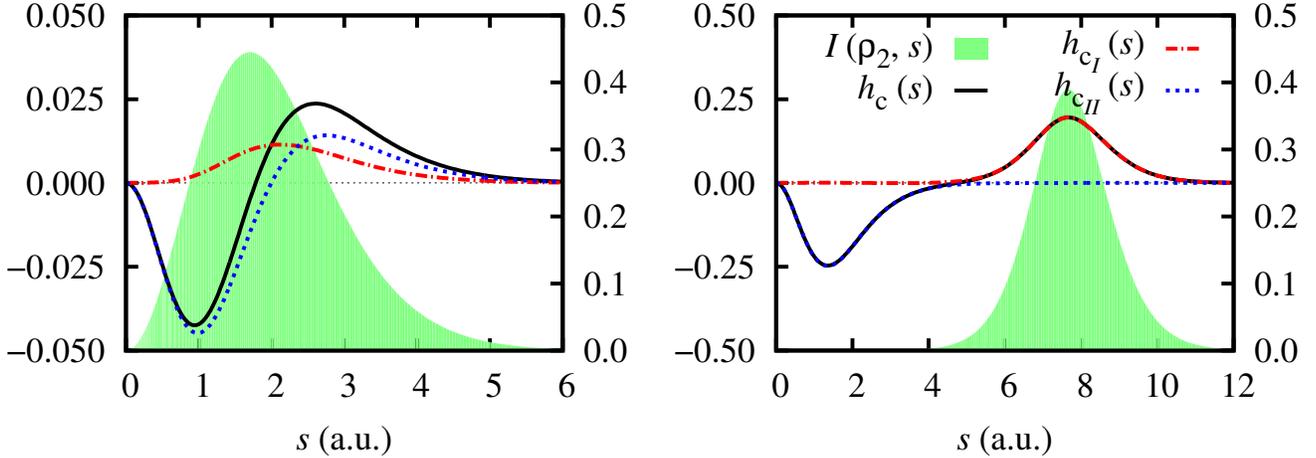}
  \caption{Coulomb hole (black), $h_{\text{c}_I} (s)$ (red) and $h_{\text{c}_{II}} (s)$ (blue) correlation components, and the intracule density (green shadowed region, right $y$-axis) of $^1\Sigma^+_u$ H$_2$ calculated with STO-3G basis set at the equilibrium distance (1.39 a.u.) and at 7.56 a.u..}
  \label{fig:h2_STO3G_hole}
\end{figure}

\begin{figure}
  \centering
  \includegraphics[width=1.0\linewidth]{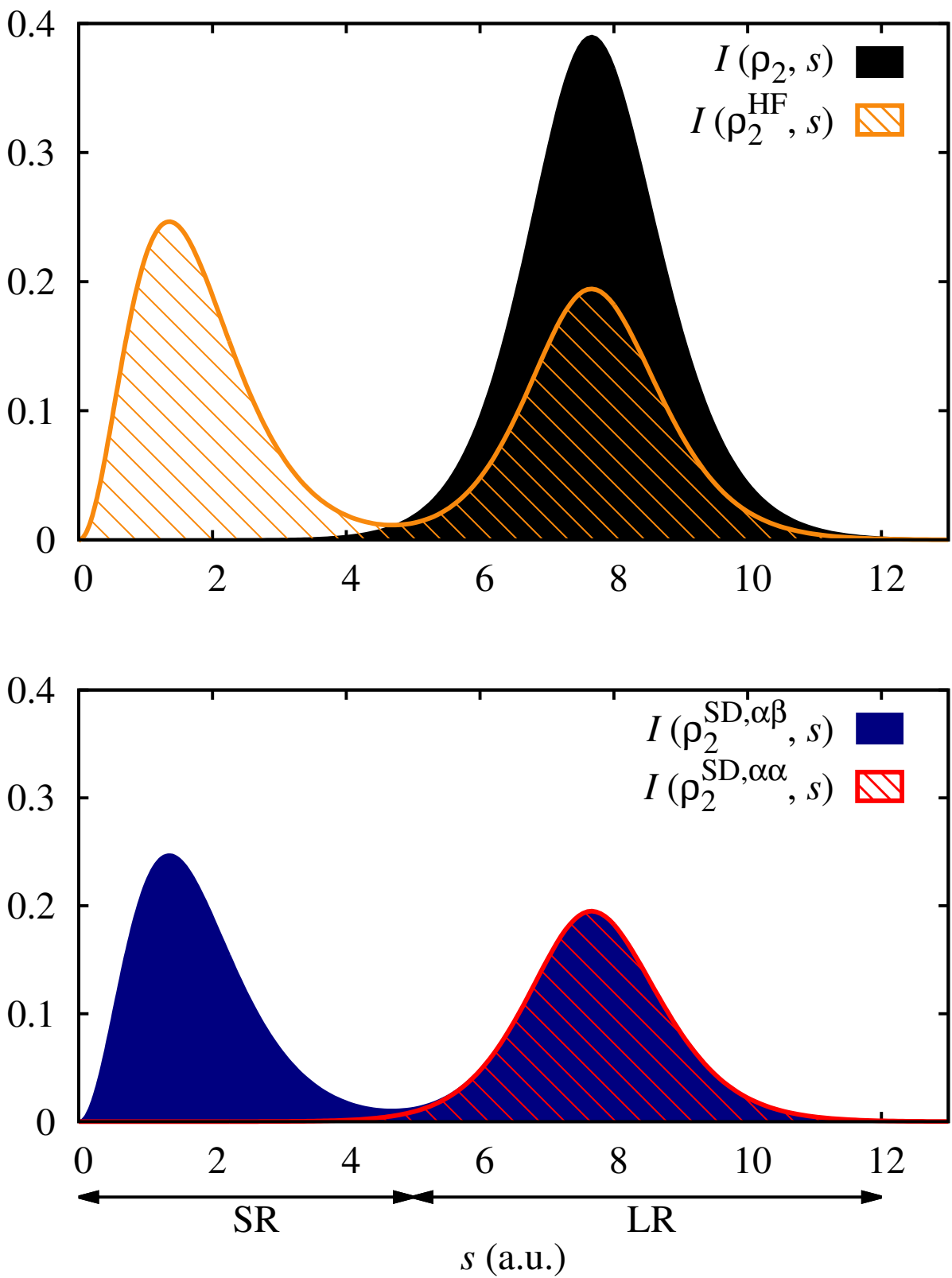}
  \caption{The intracule probability densities of H$_2$ at $R$ = 7.56 a.u. bond distance calculated with the STO-3G basis set. Top: the exact (filled black) and the Hartree--Fock (dashed orange) intracules. Bottom: the opposite- (filled blue) and same-spin (dashed red) components of the SD pair density approximation intracule. The filling has been included as a visualization aid.}
  \label{fig:h2_intracules}
\end{figure}

Due to its simplicity,
H$_2$ in a minimal basis (STO-3G)\cite{hehre:69jcp} becomes a perfect model for understanding the partition of the Coulomb hole. 
Let us consider the equilibrium and a stretched geometry. In both cases, the Coulomb hole is negative at the short range, indicating that HF overestimates the number of electron pairs at short interelectronic distances (see Figure~\ref{fig:h2_STO3G_hole}). The hole of H$_2$ at equilibrium becomes positive for values of $s$ larger than
the bond length, due to HF's underestimation of number
of electron pairs at these interelectronic distances. The $c_I$ part of the hole is positive and rather small, whereas $c_{II}$ accounts for almost the entire shape of the Coulomb hole.
The latter indicates that the role of nondynamic correlation is small at equilibrium. At the stretched geometry, the magnitude of the Coulomb hole is considerably larger, and the maximum of the hole coincides with the maximum of the intracule of the pair density. In this case, HF does a worse job describing the distribution of electron pairs, providing significant errors on both the short- and long-range components of the Coulomb hole.
As a result of the separability problem, which originates from the impossibility to separate two electrons occupying the same orbital on a restricted single-determinant wavefunction,~\cite{szabo:12book} HF
underestimates the number of electron pairs formed at large electron-electron separations and overestimates the number of pairs formed at short separations (see Figure~\ref{fig:h2_intracules}). 
Interestingly, the two components of the Coulomb hole provide a clear-cut separation of both phenomena: $c_I$ corresponds to the electron pairs missed by HF at long ranges and $c_{II}$ corrects the overestimation of electron pairing at short range. Since the Coulomb hole is the correction to HF's two-electron distribution, its energetic contribution to the total energy can be regarded as the correction to the HF electron-electron repulsion. 
In this sense, although $\text{c}_I$ contribution is the largest, both hole components contribute to the correlation repulsion energy. The $\text{c}_I$ contribution is thus small and comes entirely from the short-range
part (which is not apparent in Figure~\ref{fig:h2_STO3G_hole}, unless we plot $h_{\text{c}_I}(s)/s$) because $h_{\text{c}_I}(s)/s$ tends to zero for large $s$ as we reach the dissociation limit. 
Conversely, other properties based on the pair density, such as the covariance of the electron populations of the two H atoms,~\cite{matito:06jce} are affected by the value of $\text{c}_I$ at all ranges; for these properties, correctly retrieving $h_{\text{c}_I}(s)$ at short ranges would not be enough.\newline

\begin{figure}
  \centering
  \includegraphics[width=1.0\linewidth]{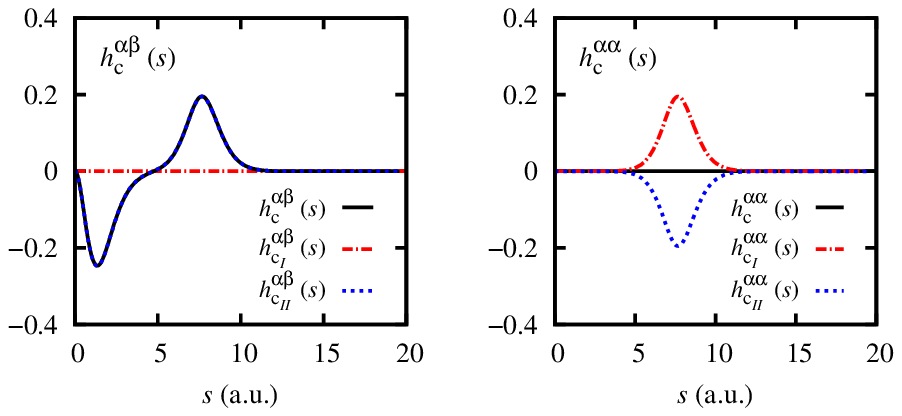}
  \caption{The opposite-spin (left) and same-spin (right) components of the FCI/STO-3G Coulomb hole ($h_{\text{c}}$), $h_{\text{c}_{I}}(s)$ (red) and $h_{\text{c}_{II}}(s)$ (blue) correlation components of H$_2$ at $R$ = 7.56 a.u..}
  \label{fig:h2_AAABhole}
\end{figure}

Let us now dissect the hole and its components into spin contributions for the stretched geometry (see Figure~\ref{fig:h2_AAABhole}). The ground state of H$_2$ has one electron of each spin; thus, the same-spin contributions of the Coulomb hole are zero, $h_{\text{c}}^{\alpha\alpha} (s) = 0$. 
However, the same-spin contribution of the correlation components of the Coulomb hole, $h^{\alpha\alpha}_{\text{c}_{\textit{I}}} (s)$ and $h^{\alpha\alpha}_{\text{c}_{\textit{II}}}(s)$, is not zero. The latter reveals that the Coulomb hole components do not have any physical meaning; they are simply mathematical objects defined for convenience. Indeed, the so-defined $h_{\text{c}_I}(s)$ of H$_2$ at dissociation arises uniquely from $h^{\alpha\alpha}_{\text{c}_{\textit{I}}} (s)$, which completely reproduces the long-range behavior of the total Coulomb hole. Conversely, $h^{\alpha\beta}_{\text{c}_{\textit{I}}} (s)$ is very small by definition, because neither HF nor SD pair densities explicitly introduce opposite-spin correlation, and the difference between
HF and the exact density is expected to be small (in the present case, where we employ a minimal basis set, the density difference is zero in the limit of H$_2$ dissociation). The long-range opposite-spin correlation is indirectly introduced in the $c_I$ component through the addition of the $h^{\alpha\alpha}_{\text{c}_{\textit{I}}} (s)$ contribution. Obviously, for systems of only one alpha electron, $h^{\alpha\alpha}_{\text{c}_{\textit{I}}} (s)=- h^{\alpha\alpha}_{\text{c}_{\textit{II}}} (s)$ and, therefore, $h^{\alpha\alpha}_{\text{c}_{\textit{II}}} (s)$ is compensating for the long-range contribution of $h^{\alpha\beta}_{\text{c}_{\textit{II}}} (s)$, making $h_{\text{c}_{\textit{II}}}(s)$ short-ranged. 
In other words, $h_{c_I}(s)$ captures the long-range component of the Coulomb hole through the intracule of   $\rho_2^{\text{SD},\alpha\alpha}(\boldsymbol{1},\boldsymbol{2})$. In particular, the latter contributes a quadratic term on $r_{12}$ at short ranges, and its long-range contribution is dominated by the $\rho^{\alpha}(\boldsymbol{1})\rho^{\alpha}(\boldsymbol{2})$ term in $\rho_2^{\text{SD},\alpha\alpha}(\boldsymbol{1},\boldsymbol{2})$. Hence, as a first approximation to the long-range part of $\Delta\rho_2^{c_I}(\boldsymbol{1},\boldsymbol{2})$ one could take $\rho^{\alpha}(\boldsymbol{1})\rho^{\alpha}(\boldsymbol{2})+\rho^{\beta}(\boldsymbol{1})\rho^{\beta}(\boldsymbol{2})$, which captures the long-range asymptotics of the Coulomb hole.

\subsection{H$_2$ triplet: dispersion interactions}

\begin{figure}
  \centering
  \includegraphics[width=1\linewidth]{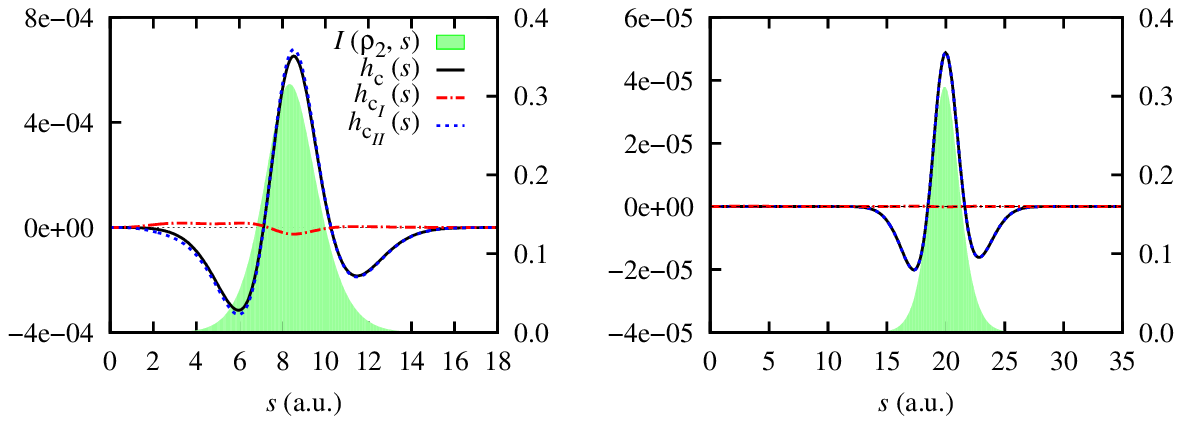}
  \caption{Coulomb hole (black), $h_{\text{c}_I} (s)$ (red) and $h_{\text{c}_{II}} (s)$ (blue) correlation components, and the intracule density (green shadowed region, right $y$-axis) of the $^3\Sigma_u^+$ state of H$_2$ molecule at 8.18 (left) and at 19.84 a.u. (right).}
  \label{fig:h2_triplet}
\end{figure}

The triplet state of H$_2$ ($^3\Sigma_u^+$) is composed solely of two electrons with the same spin and constitutes probably the simplest model to study dispersion interactions.~\cite{gritsenko:06jcp}
In contrast to the ground state singlet ($^1\Sigma_u^+$), the triplet state $^3\Sigma_u^+$ (with spin projection either $M_s = 1$ or $M_s = -1$) is qualitatively well described by the HF configuration at any interatomic distance and bears the correct dissociation.~\cite{bowman:70jcp} Indeed, compared to the singlet, the Coulomb hole is very small (see the left $y$-axis in Figure~\ref{fig:h2_triplet}). This case does not present electron entanglement, therefore, nondynamic correlation does not arise when the two fragments are separated.\newline

Long-range dispersion interactions are considered because we have employed a basis set including $p$ orbitals.~\cite{janowski:12jacs} The only peak in the Coulomb hole corresponds to dispersion interactions and, thus, it is captured by $h_{\text{c}_{\textit{II}}} (s)$.\cite{via-nadal:17pra} $h_{\text{c}_{\textit{II}}} (s)$ peaks around the interatomic separation, $R_{\text{HH}'}$,  indicating that HF underestimates the number of electron pairs at $R_{\text{HH}'}$, which are placed at shorter and longer distances.
On the other hand, there is almost no $h_{\text{c}_{\textit{I}}} (s)$ contribution at equilibrium, and it is zero at the stretched geometry. The latter is due to $\rho_2^{\text{SD}} (\boldsymbol{1},\boldsymbol{2})$ being very close to $\rho_2^{\text{HF}} (\boldsymbol{1},\boldsymbol{2})$, which results from the small changes induced by electron correlation into the density or the first-order density. As we have recently proved,~\cite{via-nadal:17pra,via-nadal:19jpcl} dispersion interactions are characterized by a universal feature:  $h_{\text{c}_{\textit{II}}}(R_{\text{HH}'})$ decays as $R_{\text{HH}'}^{-3}$ for large $R_{\text{HH}'}$.
\newline

\subsection{He-Ne atomic series}

\begin{figure}
  \centering
  \includegraphics[width=1.0\linewidth]{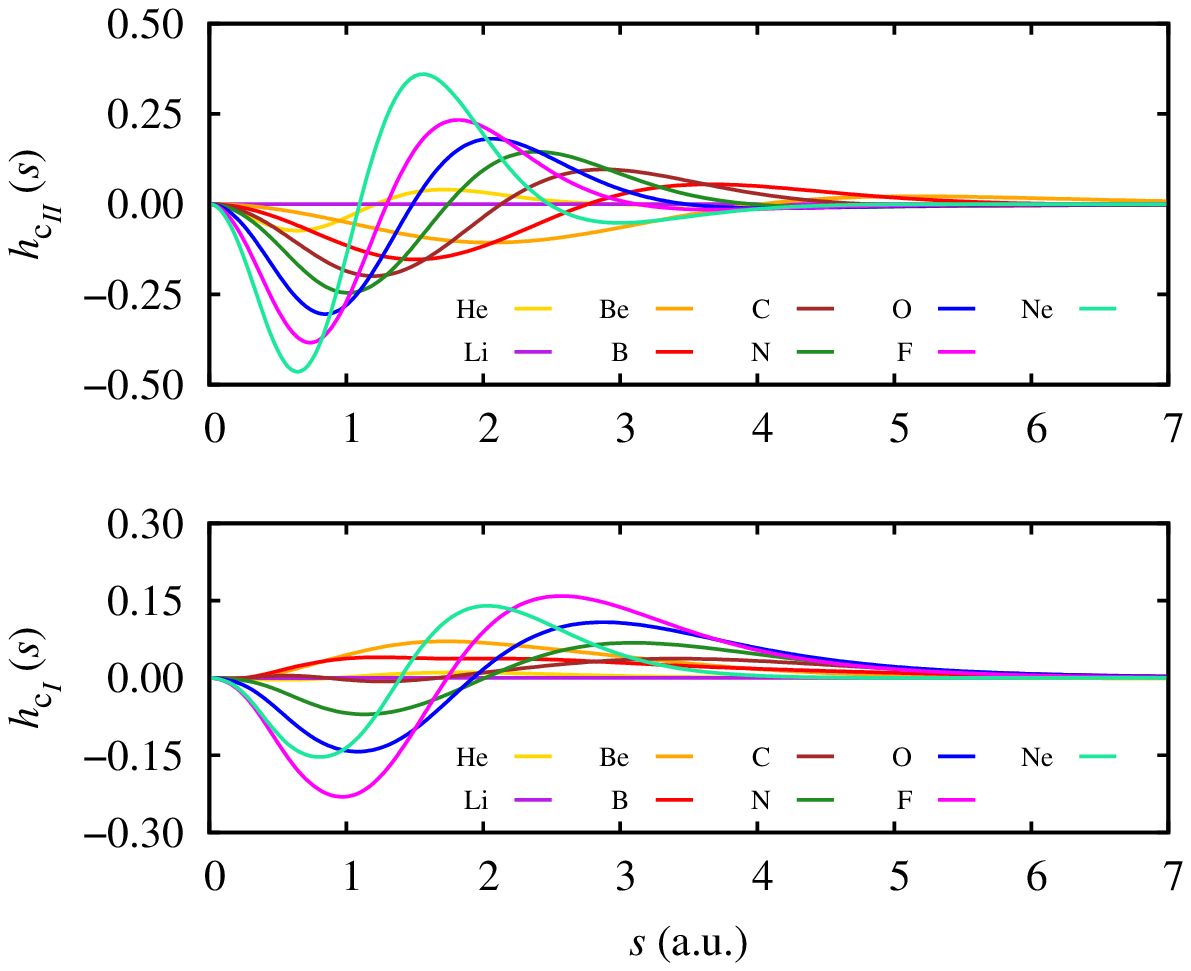}
  \caption{$h_{\text{c}_{II}}$ (top) and $h_{\text{c}_{I}}$ (bot) correlation components of the Coulomb hole of the He-Ne atomic series in their lowest-lying multiplicities.}
  \label{fig:atoms_dc}
\end{figure}

Since correlation increases with the number of electrons, the Coulomb hole is expected to increase with the atomic number, $Z$. The magnitude of the hole increases and shrinks with $Z$, whereas $\langle s \rangle$ decreases due to a larger attractive potential. However, the latter increase is not monotonic (see the plot for $h_{\text{c}}(s)$ in Figure~S4). The splitting of the hole into the two correlation components reveals that $h_{\text{c}_{II}} (s)$ presents a systematic growth with $Z$ (see the magnitude increase in the top plot of Figure \ref{fig:atoms_dc}). Conversely, $h_{\text{c}_{I}} (s)$ reveals characteristic shapes according to the nature of each atom, in agreement with the expected nondynamic correlation behavior in these systems (bottom plot in Figure \ref{fig:atoms_dc}).\newline

\begin{figure}
  \centering
  \includegraphics[width=1.0\linewidth]{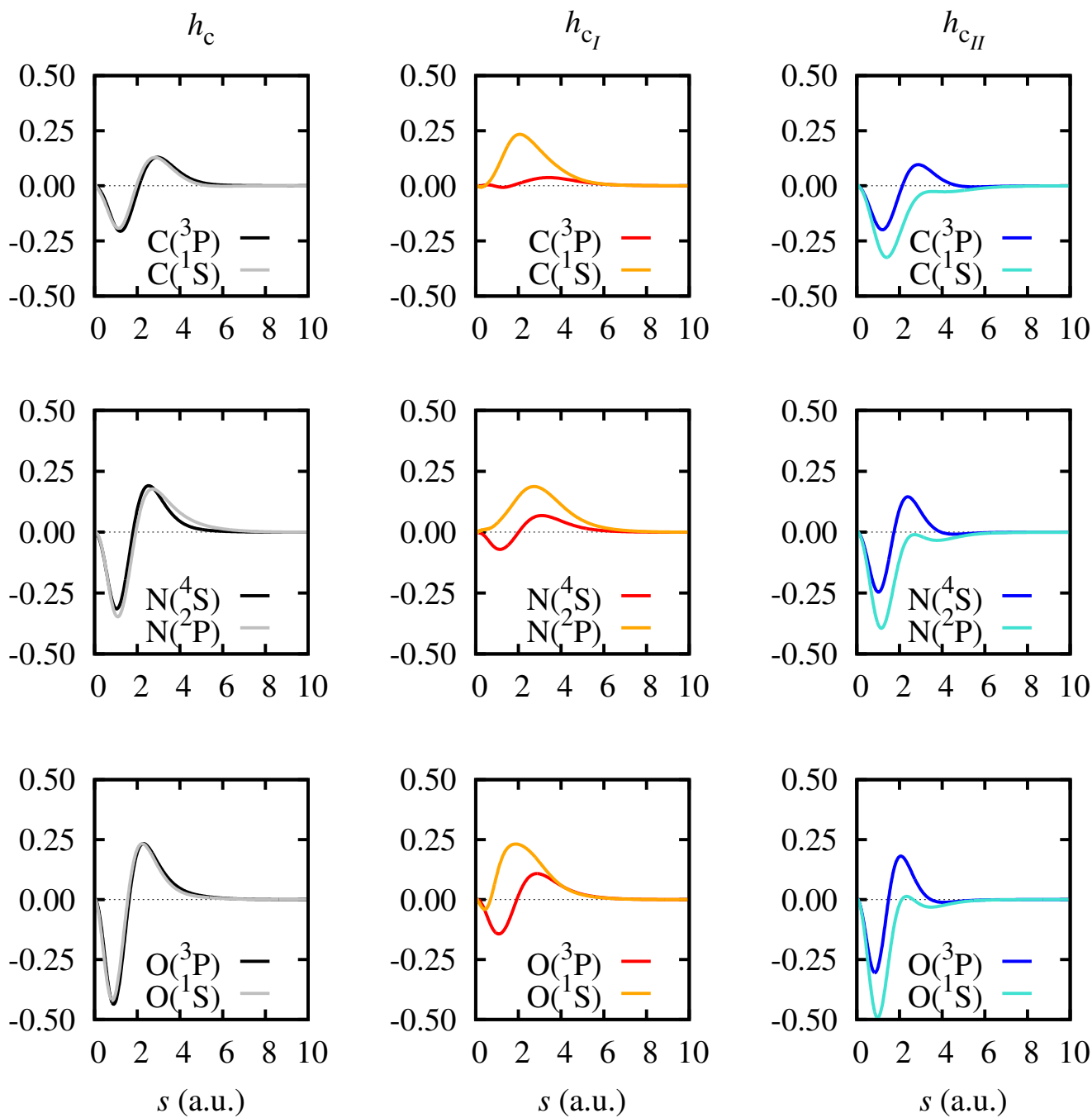}
  \caption{$h_{\text{c}} (s)$, $h_{\text{c}_I} (s)$, and $h_{\text{c}_{II}} (s)$ of carbon, nitrogen, and oxygen in their ground state (black, red and blue, respectively) and minimum multiplicity state (gray, orange and turquoise).}
  \label{fig:multiplicity_CNO}
\end{figure}

\begin{figure}
  \centering
  \includegraphics[width=1.02\linewidth]{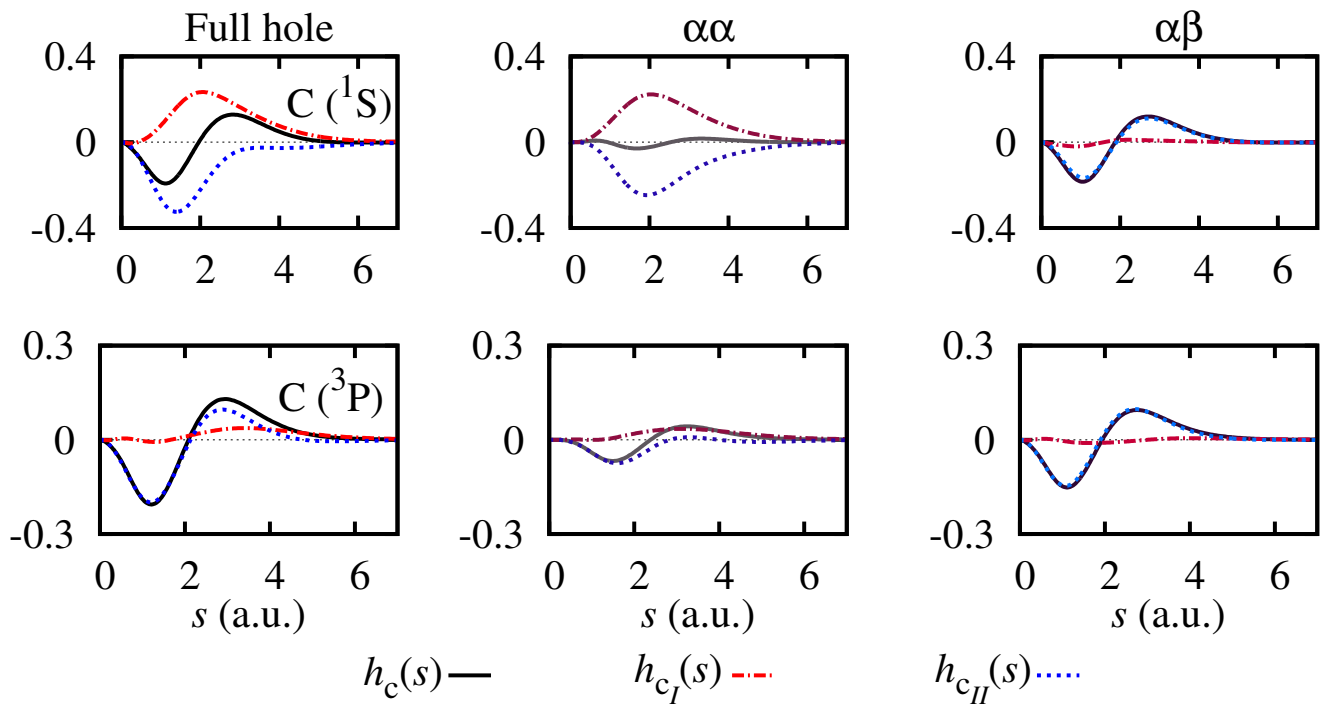}
  \caption{The Coulomb hole, $h_{\text{c}} (s)$, and its correlation components, $h_{\text{c}_I} (s)$ and $h_{\text{c}_{II}} (s)$, (left column), the $\alpha\alpha$ (middle column) and $\alpha\beta$ contributions to the Coulomb hole (right column) for singlet state carbon (top row) and the triplet, ground state (bottom row).}
  \label{fig:spinsplit_CNO}
\end{figure}

We have studied different multiplicities for carbon, nitrogen, and oxygen: their lowest-lying states (highest-spin multiplicity) and the states with the lowest multiplicity. Their holes are compiled in Figure~\ref{fig:multiplicity_CNO}. The most outstanding feature is that the Coulomb hole of all the atoms looks practically identical regardless of the multiplicity (see the first column of Figure~\ref{fig:spinsplit_CNO}) and, therefore, one expects a similar correlation contribution to the electron-electron repulsion. The nature of electron correlation is different for the various state multiplicities; therefore, we can conclude that a mere analysis of the shape of the Coulomb hole does not provide insight into the electron correlation of these systems.
Conversely, the Coulomb hole components display different shapes according to the spin state.\newline

More precisely, each state presents a characteristic profile, independently of the atom considered. In the ground states (highest-spin multiplicities), both $\text{c}_{\textit{I}}$ and $\text{c}_{\textit{II}}$ components present the conventional hole shape, being negative at short interelectronic distances and positive at large ones.  $h_{{c}_{\textit{II}}}(s)$ is, in general, larger than $h_{\text{c}_{I}}(s)$ and both increase systematically with the number of electrons (see Figure \ref{fig:multiplicity_CNO}) ---in line with the fact that the HF description loses accuracy from C to O. Conversely, the atoms in their lowest spin multiplicity present mostly positive ${h_{\text{c}_{I}}}(s)$ and short-ranged negative $h_{\text{c}_{\textit{II}}}(s)$.\newline

The reason behind these characteristic profiles is explored through the spin decomposition of the holes. We have used the carbon atom to illustrate it. As we can see in Figure~\ref{fig:spinsplit_CNO}, the spin components of the Coulomb hole present similar profiles in singlet carbon (C($^1$S)) and singlet H$_2$ (compare to Figure~\ref{fig:h2_AAABhole}). The shape of the Coulomb hole is mainly given by the opposite-spin $h_{\text{c}_{II}} (s)$ component, and the same-spin components are more important in the description of the singlet than in the triplet. The latter feature can be explained from the nature of $\rho_2^{\alpha\alpha,\text{SD}}(\boldsymbol{1},\boldsymbol{2})$, which contributes to $h^{\alpha\alpha}_{\text{c}_I} (s)$. Since $h^{\alpha\alpha}_{\text{c}_I} (s)$ is greater than zero, its magnitude can be related to the inability of $\rho_2^{\alpha\alpha,\text{SD}}(\boldsymbol{1},\boldsymbol{2})$ to reproduce the number of electron pairs of the same spin upon integration over $\boldsymbol{1}$ and $\boldsymbol{2}$. The more considerable the electron correlation effects on the first-order density, the larger the deviation from the number of electron pairs of the same spin.\cite{matito:07fd} Therefore, singlet carbon, which is much more affected by nondynamic correlation, presents larger values of $h^{\alpha\alpha}_{\text{c}_I} (s)$ and, since the total Coulomb hole is mainly given by the opposite-spin component, $h^{\alpha\alpha}_{\text{c}_{II}}(s)$ will be large and (partially) compensate for $h^{\alpha\alpha}_{\text{c}_I} (s)$ (see Figure~\ref{fig:spinsplit_CNO}).
Table~\ref{table_spinsplit} shows that the trace deviation of the like-spin SD pair density is larger in the singlet than in the triplet.  

\begin{table}
\begin{center}
\begin{tabular}{llll}
 & Ne & C($^3$P) &C($^1$S) \\
\hline\hline
Tr$\left[ \rho_2^{\alpha\alpha} \left( \boldsymbol{1},\boldsymbol{2} \right) \right]$  & 20.00    & 7.00    & 6.00  \\
\\
Tr$\left[ \rho_2^{\alpha\alpha,\text{SD}}  \left( \boldsymbol{1},\boldsymbol{2} \right) \right]$        & 20.06 &  7.11 & 6.59 \\
\hline
\end{tabular}
\caption{Same-spin pair-density traces for Ne, the singlet and triplet states of carbon atom. Tr$\left[ \rho_2^{\alpha\alpha}  \left( \boldsymbol{1},\boldsymbol{2} \right) \right]=N^{\alpha}\left(N^{\alpha}-1\right)$.}
\label{table_spinsplit}
\end{center}
\end{table}

\subsection{The Be$_2$ Coulomb hole}

The description of both Be atom and its dimer is very sensitive to the level of theory and the quality of the basis set employed\cite{NIST} (see also Supporting Information, where we follow a previous strategy~\cite{matito:10pccp,cioslowski:11jctc,rodriguez:19co} to analyze the effect of the basis set). 
Due to the well-known near-degeneracy of $2s$ and $2p$ orbitals in Be, the binding of both fragments and, in general, the potential energy curve (PEC) of Be$_2$ has been the subject of study of both experimentalists and computational chemists alike. One of the most intriguing features of the PEC of Be$_2$ is a change of slope that generates two potential minima, one corresponding to the equilibrium geometry at $4.72$ a.u., and the second one at twice the equilibrium distance. Because of this double well, a single-reference method with a basis set that includes neither $d$ nor $f$ functions usually gives the second geometry as the equilibrium one, or even a repulsive, non-binding curve. \cite{schmidt:10jpca,elkhatib:14jpca,noga:92cpl} In our study, we have run a frozen-core full configuration interaction (FCI) calculation with the aug-cc-pVTZ basis set to study the Coulomb hole. It has been shown elsewhere~\cite{elkhatib:14jpca} that the core electrons do not play an important role in the electronic description of Be$_2$ because the triple and quadruple excitations of the valence electrons are the ones responsible for most electron correlation. \newline

\begin{figure}
  \centering
  \includegraphics[width=1.0\linewidth]{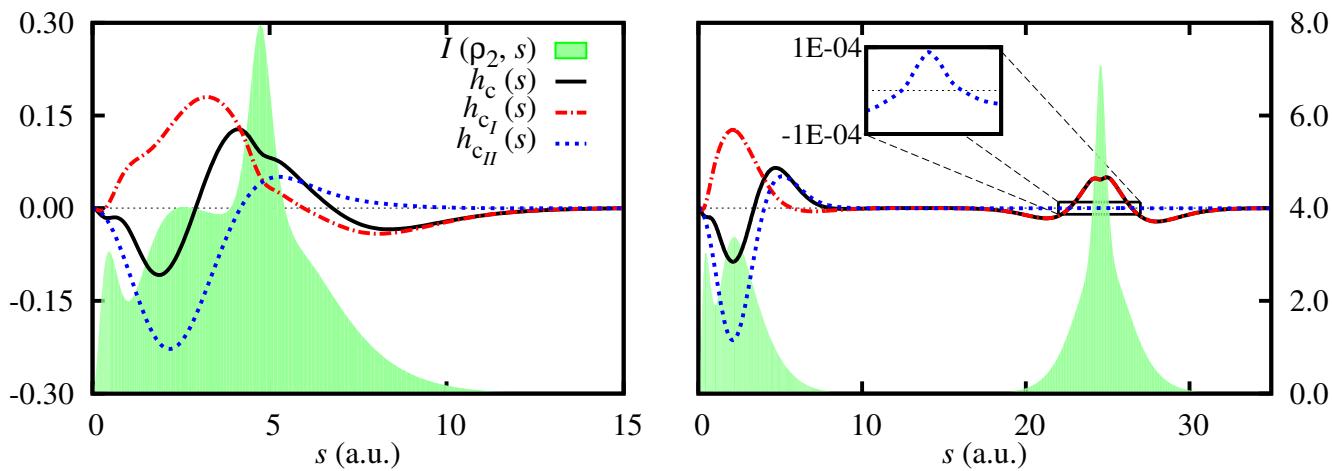}
  \caption{Coulomb hole (black), $h_{\text{c}_I} (s)$ (red) and $h_{\text{c}_{II}} (s)$ (blue) correlation components, and the intracule density (green shadowed region, right $y$-axis) of the Be$_2$ molecule at 4.72 a.u. (left) and 24.57 a.u. (right) bond lenghts.}
  \label{fig:be2_holes}
\end{figure}

Figure~\ref{fig:be2_holes} contains the Coulomb holes of two geometries of Be$_2$. Two interesting features of $h_{\text{c}} (s)$ are the presence of a small \textit{bump} around the hole evaluated at the interatomic distance, $h_{\text{c}} (s = R)$, and a shoulder at the small $s$ region for both geometries. The latter is connected with the lack of correlation of the core electrons (see the magnification in Figure~S2 and Ref.~\citenum{rodriguez:19co} for a discussion on the effect of core electrons in the Coulomb hole).
For both geometries, $h_{\text{c}_{I}} (s)$ takes large numbers due to the nondynamic correlation arising from the near-degeneracy of the $2s$ and $2p$ orbitals of the beryllium atom; this multireference character is preserved in the dimer. It has actually been reported that the valence molecular orbitals $3\sigma_g$ and $3\sigma_u$, which arise from linear combinations of $2s$ and $2p$, show relatively large orbital occupation numbers and, hence, should be considered in the calculation.\cite{elkhatib:14jpca,noga:92cpl} Indeed, the FCI occupation number of both $3\sigma_g$ and $3\sigma_u$ is 0.0589 at $R =$ 24.57 a.u.. Conversely, since the $2\sigma$ orbitals are occupied in the HF description, it cannot provide an accurate 1-RDM at this stretched geometry and, therefore, $\Delta\rho_2^{\text{c}_{I}} (\boldsymbol{1},\boldsymbol{2})$ is large. Certainly, the wrong dissociation provided by HF and the strong multireference character of the dissociated Be fragments make the HF 1-RDM a very poor reference. The long-range part of the Coulomb hole of stretched Be$_2$ presents a maximum that belongs to $h_{\text{c}_{\textit{I}}} (s)$, caused by the presence of the valence electrons localized near the respective nuclei.  The long-range part of $h_{\text{c}_{\textit{II}}} \left(s\right)$ in stretched Be$_2$ presents a minuscule yet positive area due to dispersion interactions.\newline

The most remarkable feature of the Coulomb hole splitting of this molecule is the dominance of both types of correlation at short range. In most of the cases studied thus far, the short-range part of the Coulomb hole has always been dominated by $h_{\text{c}_{\textit{II}}} (s)$, which defined the shape of the short-range part of the total hole $h_{\text{c}} (s)$. At the equilibrium and stretched geometries, the short-range part of the total hole of Be$_2$ is molded by both correlation contributions. \newline

\subsection{Types A and B of nondynamic correlation: Be(Z) and H$_2$}

Hollett and Gill recognized two types of nondynamic correlation that are classified according to the ability of the unrestricted Hartree--Fock (UHF) method to capture them. \cite{hollett:11jcp} The first type, type A, arises from absolute near degeneracies, for example, those that occur when a homolytic dissociation takes place. As the interatomic separation distance $R$ increases, the energy gap ($\epsilon_{gap}$) between the highest occupied and lowest unoccupied molecular orbitals (HOMO and LUMO, respectively) becomes smaller until it vanishes and these orbitals become degenerate. Certainly, RHF cannot describe the localization of the electrons at each nucleus, whereas the unrestricted formalism succeeds in doing so by getting rid of the ionic description. The second type, named type B, arises in the presence of relative near-degeneracies. The HOMO-LUMO gap widens in the Be isoelectronic series as the effective charge $Z$ increases.  However, the difference between gap increments, $\Delta \epsilon_{gap}$, remains constant. UHF cannot portray the nondynamic correlation in this scenario, and multireference methods are required for an accurate description. Identifying correlation types A and B is a challenging test for correlation indicators.\cite{ramos-cordoba:16pccp} The present section studies whether correlation types A and B can be detected by $h_{\text{c}_{\textit{I}}} (s)$. Namely, we analyze a typical case of type B correlation, the isoelectronic series of Be, Be($Z$) with $Z$ = 3--8, and the dissociated and equilibrium geometries of H$_2$ as an example of type A correlation. \newline

Many studies on the consistency and convergence of basis sets using the Be atom exist in the literature, reflecting the difficulty to correctly reproduce the short-range interactions in this system. \cite{galvez:03cpl,schmidt:10jpca} We have optimized an even-tempered basis set of 10 \textit{s, p}, and \textit{d} functions to perform the calculations of Be($Z$). Information about the optimization is provided in the Supporting Information. \newline

It has been demonstrated elsewhere that the correlation energy in Be grows linearly with $Z$ due to the $2s-2p$ near-degeneracy.\cite{chakravorty:93pra} Consequently, the energy difference between UHF and RHF, $\Delta E$, is expected to increase with $Z$. Instead, Hollet and Gill demonstrated the contrary in their work: the molecules present a triplet instability and UHF can describe nondynamic correlation from $Z$ = 3.0 to $Z$ = 4.25, yielding lower total energies with respect to RHF. Conversely, the fraction of energy recovered by UHF decreases as $Z$ increases, and the UHF description in $5 \leq Z \leq 8$ is equivalent to the RHF one (see Table \ref{table_uhf_beiso}). The HOMO-LUMO gap widens with the effective charge $Z$, but the difference between gaps, $\Delta \epsilon_{\text{gap}}$, remains constant, indicating a linear gap growth. 
\begin{table*}
\begin{center}
\begin{tabular}{l|llllllll}
$Z$ & $E_{\text{RHF}}$ & $E_{\text{UHF}}$ & $\Delta E$ & $\langle S^2 \rangle_{\text{UHF}}$ & $\epsilon_{\text{HOMO}}$  & $\epsilon_{\text{LUMO}}$ & $\epsilon_{\text{gap}}$ & $\Delta\epsilon_{\text{gap}}$ \\
\hline\hline
3 & -7.38012 & -7.39086 & -0.01074 & 0.7132 & 0.04521 & 0.02639 & 0.21864 & 0.15874 \\
4 & -14.57022 & -14.57059 & -0.00036 & 0.1378 &-0.31300 & 0.06438 & 0.37738 & 0.22385 \\
5 & -24.23339 & -24.23339 & 0.00000                 & 0.0000   &-0.87412 & -0.27289 & 0.60123 & 0.22636 \\
6 & -36.39601 & -36.39601 & 0.00000                 & 0.0000   &-1.69410 & -0.86651 & 0.82759 & 0.22368 \\
7 & -57.07217 & -57.07217 & 0.00000                 & 0.0000   &-2.76685 & -1.71558 & 1.05127 & 0.21915 \\
8 & -68.22939 & -68.22939 & 0.00000                 & 0.0000   &-4.08835 & -2.81793 & 1.27042 & - \\
\hline
\end{tabular}
\caption{The Restricted Hartree--Fock energy, Unrestricted Hartree--Fock energy, their energy difference, the spin contamination from the UHF calculation, the HOMO and LUMO energies and their difference, and the difference between energy gaps for Be-like ions with $3 \leq Z \leq 8$.}
\label{table_uhf_beiso}
\end{center}
\end{table*}

\begin{figure}
  \centering
  \includegraphics[width=1.0\linewidth]{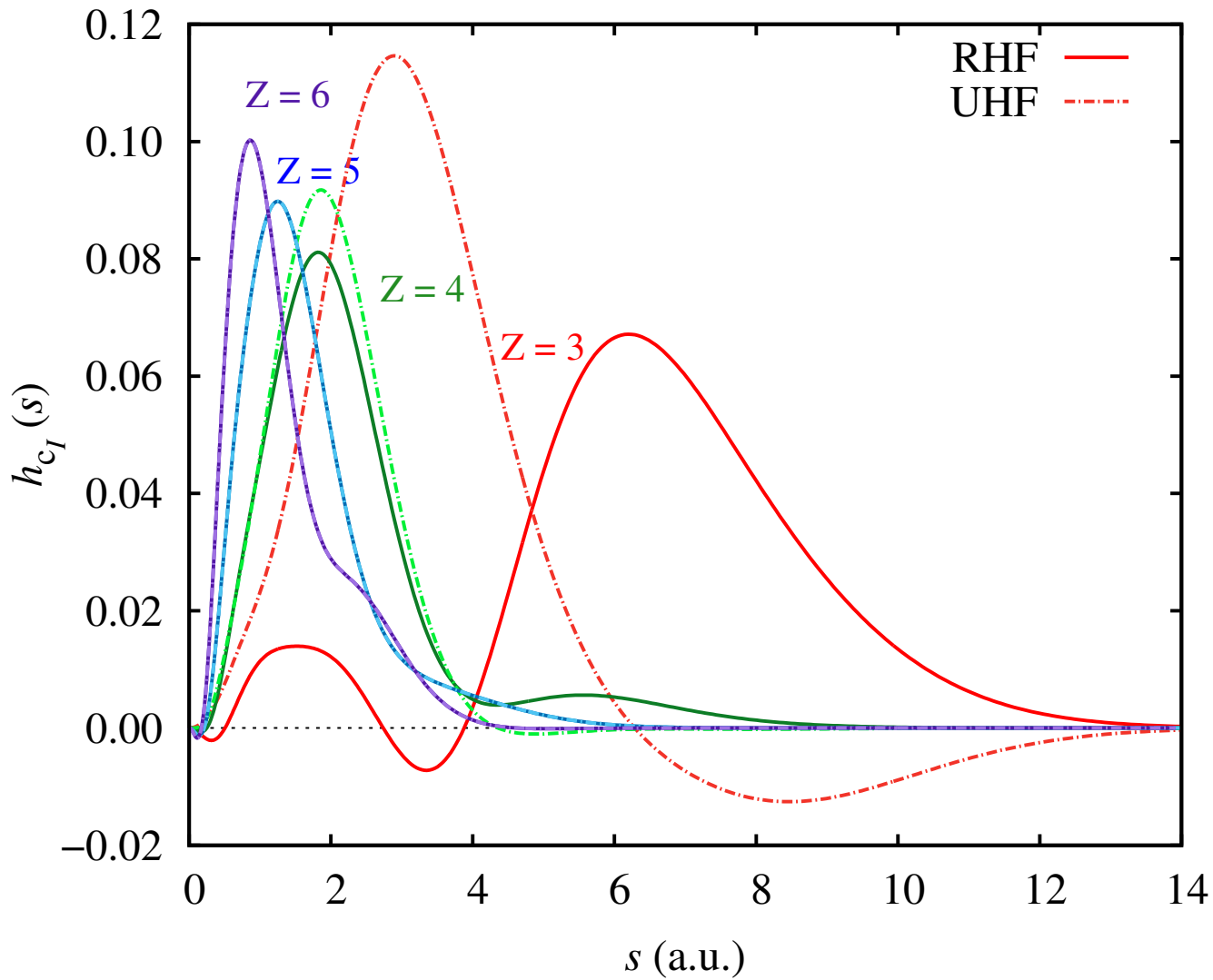}
  \caption{$h_{\text{c}_{I}} (s)$ component of the Coulomb hole of Be-like atoms, Be($Z$), with $Z$ = 3, 4, 5 and 6. Solid (dashed) lines depict the holes described using as reference the restricted (unrestricted) formalism of Hartree--Fock, that is, $h_{\text{c}_{I}} (s) = I\left(\rho_2^{\text{SD}},s\right) - I\left(\rho_2^{\text{RHF}},s\right)$ for solid lines, and $h_{\text{c}_{I}} (s) = I\left(\rho_2^{\text{SD}},s\right) - I\left(\rho_2^{\text{UHF}},s\right)$ for dashed lines.}
  \label{fig:bz_ndc}
\end{figure}

Figure~\ref{fig:bz_ndc} contains the $h_{\text{c}_{I}} (s)$ contribution for Be-like ions with $3 \leq Z \leq 6$. The numbers displayed in Table~\ref{table_uhf_beiso} and Ref.~\citenum{hollett:11jcp} show that a broken symmetry solution does not exist for $Z$ = 5 and 6 and, hence, the UHF and RHF holes are indistinguishable. The UHF and RHF holes for the Be atom are quite similar, whereas for $Z$ = 3 both descriptions are no longer equivalent.

Figure~S3 of the Supporting Information provides the Coulomb hole decomposition of a UHF calculation of the H$_2$ with a minimal basis set. The latter is barely indistinguishable from the FCI counterpart (Figure~\ref{fig:h2_STO3G_hole}), indicating that $h_{\text{c}_{I}} (s)$ also accounts for type A correlation. In fact, we can easily prove analytically that the FCI and UHF pair density and the 1-RDM are identical in a minimal basis set. From the examples in this section, we conclude that $h_{\text{c}_{I}} (s)$ can describe nondynamic correlation, regardless of its type.

\subsection{C$_{3v}$-H$_4$}

\begin{figure}
  \centering
  \includegraphics[width=0.60\linewidth]{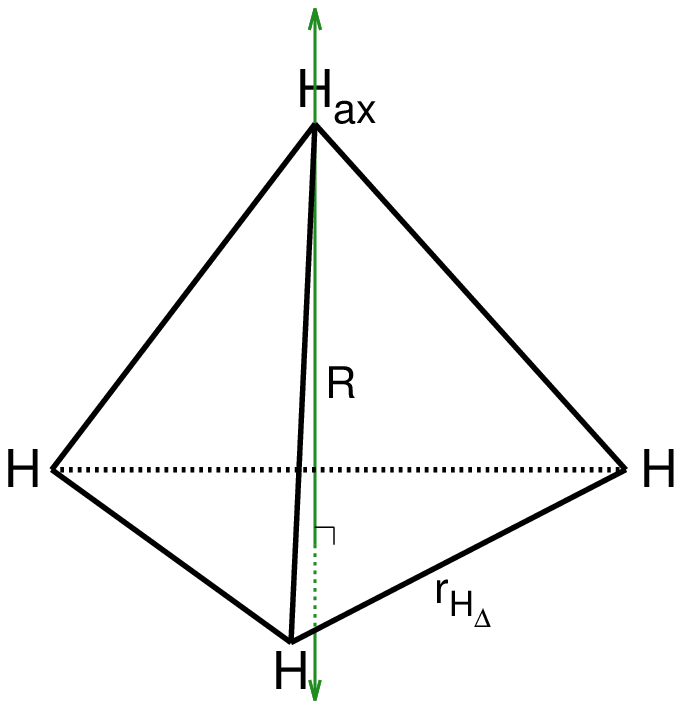}
  \caption{Graphical representation of H$_4$ at the C$_{3v}$ symmetry point group, where the position of the axial hydrogen (H$_{\text{ax}}$) through the axial axis $R$ (marked as green) represents the PEC analyzed. We keep r$_{H_{\Delta}}$ fixed at 1.77 a.u.. $R$ = 0 when H$_{\text{ax}}$ is at the equatorial plane.}
  \label{fig:tetrahedre}
\end{figure}

We consider the three lowest-lying states of the C$_{3v}$ PEC of H$_4$ constructed by displacing one H (hereafter, the axial H or H$_{\text{ax}}$) in the direction perpendicular to the equatorial plane where the rest of H atoms are located (see Figure~\ref{fig:tetrahedre}).~\cite{boothroyd:02jcp,aguado:94jcp,montgomery:87jcp,evleth:88jcp,jorge:93jbcs,theodorakopoulos:87tch,nicolaides:84jcp} The distance between the equatorial hydrogens ($R_{\text{H}_{{\Delta}}}$) is kept fixed at 1.77 a.u..~\cite{jorge:93jbcs}\newline

\begin{figure}
  \centering
  \includegraphics[width=1.0\linewidth]{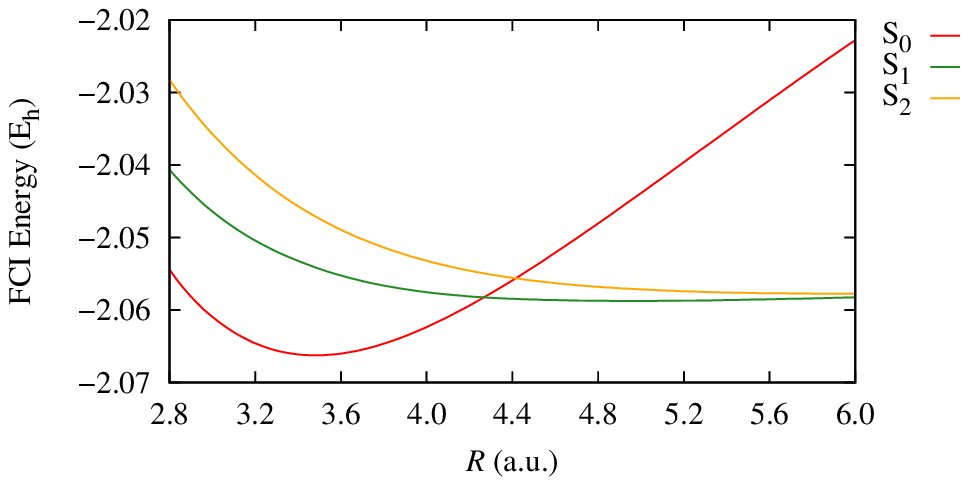}
  \caption{C$_{3v}$ PECs of H$_4$, where the coordinate axis represents the separation of the axial hydrogen H$_{\text{ax}}$ from the plane. A crossing from S$_0$ to S$_1$ takes place at 4.27 a.u.}
  \label{fig:h4c3v_pec}
\end{figure}

The FCI PEC presents an intersystem crossing around $R$ = 4.50 a.u. from S$_0$ to state S$_1$ (see Figure~\ref{fig:h4c3v_pec}). At the S$_0$ state, the system is described as two interacting H$_2$ molecules. The increase of $R$ causes one of these molecules to dissociate, leaving H$_3^{+}$ in the equatorial plane and an H$^{-}$, the dissociated hydride (H$_{\text{ax}}$). The HF description of $S_0$ at the ground state is dominant in the CI vector, and the largest occupation numbers of the natural orbitals are 1.957 and 1.936. Conversely, at $R$ = 28.35 a.u. (S$_1$), the CI vector comprises at least four highly contributing configurations, and the natural occupations are 1.954, 1.000, 0.992, and 1.000, the latter  corresponding to H$_{\text{ax}}$. \newline

\begin{figure}
  \centering
  \includegraphics[width=1.00\linewidth]{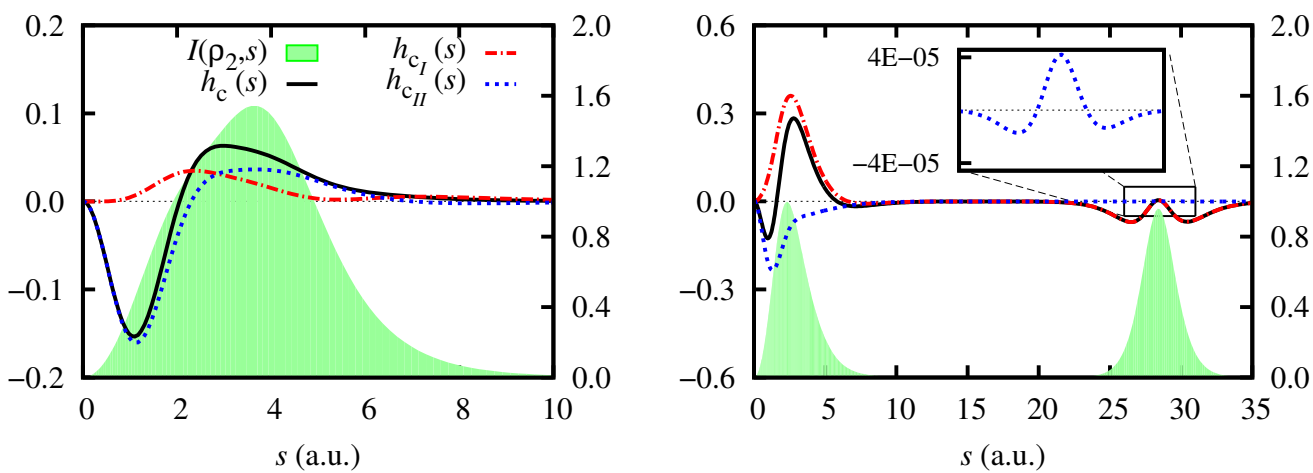}
  \caption{Coulomb hole (black), $h_{\text{c}_I} (s)$ (red) and $h_{\text{c}_{II}} (s)$ (blue) correlation components, and the intracule density (green shadowed region, right $y$-axis) of H$_4$ molecule at the C$_{3v}$ symmetry point group with fixed r$_{\text{H}_{\Delta}}$. H$_{\text{ax}}$ is placed at (left) $R$ = 3.50 a.u. (right) and $R$ = 28.35 a.u. from the equatorial plane (see Figure~\ref{fig:tetrahedre} for details).}
  \label{fig:h4c3v_holes}
\end{figure}

The magnitudes of the Coulomb holes at the equilibrium and dissociated geometries reflect the changes in the CI vector, where the S$_1$ state is more correlated than S$_0$. The correlation contributions depicted in Figure~\ref{fig:h4c3v_holes} show that $h_{\text{c}_{\textit{II}}} (s)$ is prevalent at the equilibrium geometry and defines the shape of the Coulomb hole. At the stretched geometry, however, the short-range part of the Coulomb hole is dominated by $h_{\text{c}_{\textit{I}}} (s)$, mainly caused by the multireference description of the S$_1$ state that results from the spin frustration of the three hydrogens that remain on the same plane. $h_{\text{c}_{\textit{II}}} (s)$ retains a simple shape and presents a long-range peak caused by the dispersion interaction between H$_{\text{ax}}$ and the rest of the hydrogens in the plane. $h_{\text{c}_{\textit{I}}} (s)$ at the stretched geometry presents two features we have not seen thus far in this component of the Coulomb hole: it has a substantial value at short ranges and shows important negative values at long ranges. These unique features of $h_{\text{c}_{\textit{I}}} (s)$ put forward the very deficient description of HF, which does not dissociate into fragments with an integer number of electrons. If HF dissociated into the correct number of electrons for each fragment, the shape of $h_{\text{c}_{\textit{I}}} (s)$ would not show these features (compare to the Coulomb hole in Figure~S6, which uses UHF as a reference).
The positive short-range part of $h_{\text{c}_{\textit{I}}} (s)$ reflects the excess of electrons that HF locates in the H$_3$ plane. HF also locates less than one electron in H$_{\text{ax}}$, resulting in the negative long-range part of $h_{\text{c}_{\textit{I}}} (s)$. 

\subsection{LiH and the harpoon mechanism}

\begin{figure}
  \centering
  \includegraphics[width=1.0\linewidth]{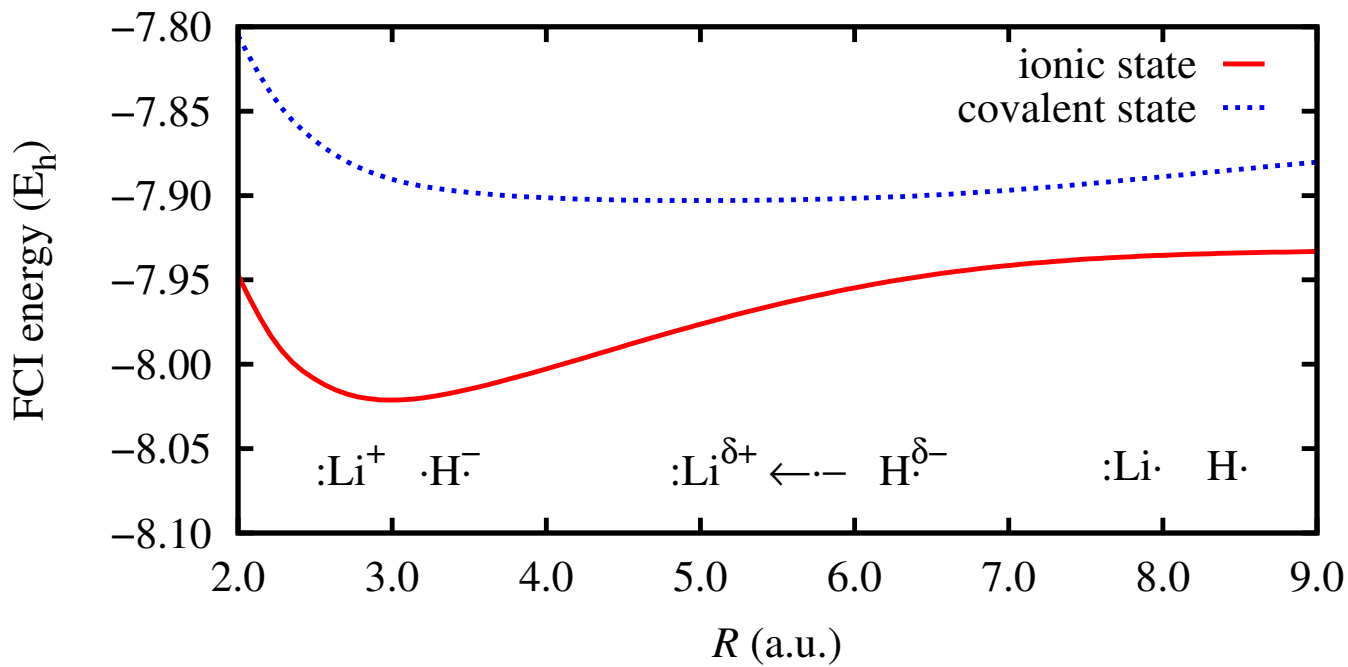}
  \caption{PECs of the lowest-lying diabatic states in LiH. An avoided crossing takes places around 6.8 a.u. Below, the harpoon mechanism is depicted.}
  \label{fig:lih_harpoon}
\end{figure}

LiH has an ionic ground state and its lowest-lying excited state is covalent.~\cite{rodriguez:16mp} Thus, in the adiabatic representation, the ground state of LiH presents an ionic bonding at equilibrium, with Li$^{+}$ and H$^{-}$, whereas the character of the state changes from ionic to covalent as the molecule dissociates (see Figure~\ref{fig:lih_harpoon}). The mechanism depicting the electron transition from hydrogen to lithium is known as the harpoon mechanism,\cite{polanyi:95acr} and it is caused by the small ionization potential of lithium and the large electron affinity of hydrogen. The potential energy curve presents an avoided crossing where the transition from the covalent state to the ionic description occurs ($R_{Li-H}=6.8$).~\cite{rodriguez:16mp}

\begin{figure}
  \centering
  \includegraphics[width=1.00\linewidth]{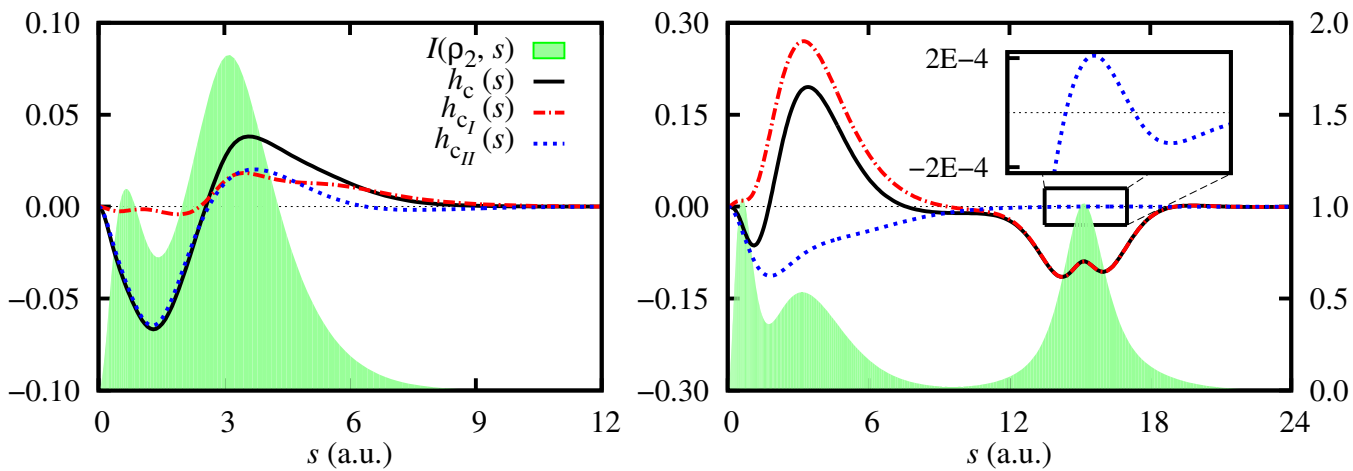}
  \caption{Coulomb hole (black), $h_{\text{c}_I} (s)$ (red) and $h_{\text{c}_{II}} (s)$ (blue) correlation components, and the intracule density (green shadowed region, right $y$-axis) of LiH at the equilibrium geometry (3.02 a.u.), and at 15.12 a.u..}
  \label{fig:lih_holes}
\end{figure}

Two different profiles, which resemble the C$_{3v}$-H$_4$ profiles depicted in Figure~\ref{fig:h4c3v_holes}, are obtained at different distances (see Figure~\ref{fig:lih_holes}).  Although HF correctly describes the ionic equilibrium geometry, the covalent character of the state at large interatomic distances is not well represented. 
Therefore, the Coulomb hole at equilibrium does not differ qualitatively from the hole of other molecules, such as H$_2$.
On the other hand, $h_{\text{c}_{\textit{I}}} (s)$ is predominant along the interelectronic distance coordinate for the stretched geometry. $h_{\text{c}_{\textit{II}}} (s)$ displays  moderate values at short range and is inappreciable at long range for the stretched LiH. As we can see in the inset plot, $h_{\text{c}_{\textit{II}}} (s)$ is characterized by a maximum, which features the universal signature of dispersion. HF also dissociates LiH into fragments with a non-integer number of electrons; therefore, the resemblance between this hole and the one of H$_4$. In fact, at the stretched geometry, the difference between the exact intracular and the UHF one is negligible (see Figure~S5), indicating that all the features of the Coulomb hole in Figure~\ref{fig:lih_holes} are entirely due to the wrong dissociation of HF.~\cite{via-nadal:17pra,via-nadal:19jpcl} \newline

\section{Conclusions}

We have studied a correlation decomposition scheme that provides a natural separation of the pair density.
The Coulomb hole's range separation into two components, $h_{\text{c}_I}(s)$ and $h_{\text{c}_{II}}(s)$, arises after 
integrating the extracule coordinate. The c$_{\textit{I}}$ component describes nondynamic long-range correlation interactions, whereas the c$_{\textit{II}}$ component features dynamic and nondynamic short-range interactions.
These components are exhaustively analyzed throughout various examples that put forward the most relevant features of this approach.\newline

First of all, through the asymptotic properties of the hole parts, we explain how the component based only on the first-order reduced density matrix (c$_{\textit{I}}$) can retrieve the long-range part of electron correlation. Second, we perform an exhaustive analysis of the hydrogen molecule in a minimal basis set, dissecting the hole contributions into spin components. Third, we analyze the simplest molecule presenting a dispersion interaction, triplet H$_2$, and how $h_{\text{c}_{II}}(s)$ helps identify it. This dispersion signature is also present in all the other molecules studied in this work, highlighting its universal character.
We also analyze the Coulomb holes of several atoms in different spin states, finding that the hole components distinguish correlation regimes that are not apparent from the entire hole.
Indeed, atoms with different spin states present the same Coulomb hole profile but its c$_{\textit{I}}$ component gives away the true multireference character of the singlet state of carbon and nitrogen atoms.\newline 

Finally, we analyze the two types of nondynamic correlation, types A and B,\cite{hollett:11jcp} and show that c$_{\textit{I}}$ can capture them. Profiles of $h_{\text{c}_{\textit{I}}} (s)$ calculated with an unrestricted reference differ from profiles calculated with a restricted wavefunction when type B nondynamic correlation is present. The more important the type B correlation, the more significant the difference between both holes. Interestingly, the correlation indicators based on natural orbitals that were also developed from this partition\cite{ramos-cordoba:16pccp} could not make such a distinction.\newline

The results of this work hold the promise to aid in developing new electronic structure methods that efficiently capture electron correlation. In particular, the models of the $c_{II}$ component can be combined with other results available in the literature\cite{rodriguez:17pccp2,cioslowski:15jcp,ramos-cordoba:14jcp} to develop new approximations in the reduced density matrix functional theory,\cite{piris:14ijqc,pernal:15tcc} although the present work is not limited to this theory.

\section*{Conflicts of interest}
There are no conflicts to declare.

\section*{Acknowledgements}
The authors thank Prof. Manuel Y\'añez and Prof. Paul W. Ayers for proposing the study of Be$_2$ and  H$_4$ at the C$_{3v}$ symmetry point group, respectively.
Grants PGC2018-098212-B-C21, BES-2015-072734, and IJCI-2017-34658 funded by MCIN/AEI/ 10.13039/501100011033 and “FEDER Una manera de hacer Europa”, and the grants funded by Diputaci\'on Foral de Gipuzkoa (2019-CIEN-000092-01), and Gobierno Vasco (IT1254-19 and PIBA19-0004) are acknowledged.

\balance

\balance

\renewcommand\refname{References}

\bibliography{gen} %
\bibliographystyle{rsc} %

\end{document}


\section{Basis set consistency in full configuration interactions calculations of Be atom}

\begin{figure}[h!]
  \centering
  \includegraphics[width=0.80\linewidth]{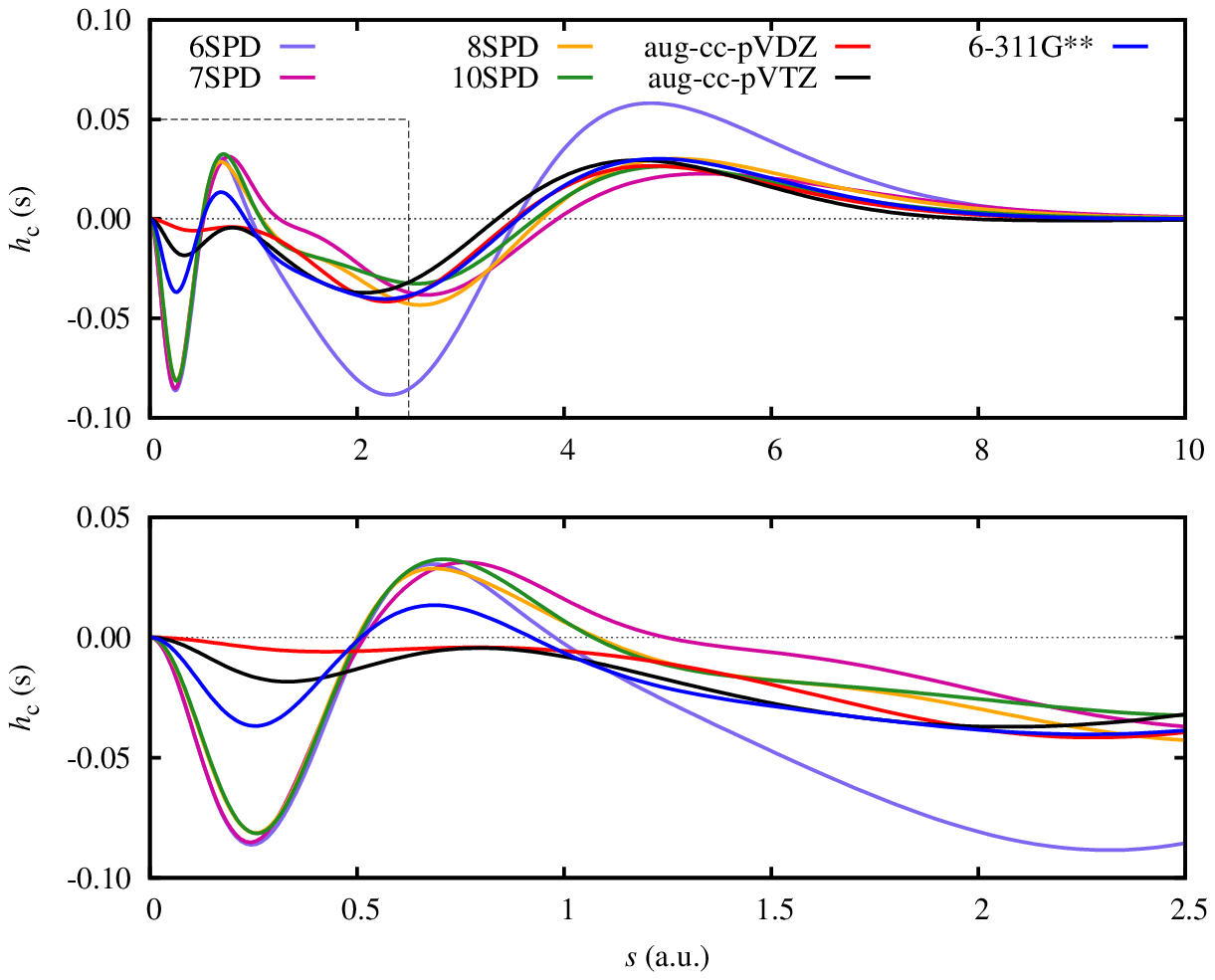} 
  \caption{Coulomb hole $h_{\text{c}} (s)$ of Be computed with different basis sets. The graph below depicts a zoomed section from the plot above, marked with dashed lines.}
  \label{fig:bz_convergence}
\end{figure}

In previous studies of the Be atom and its dimer,~\cite{galvez:03cpl,schmidt:10jpca,rodriguez:19co} it has been found that the description of the species is very sensitive to the basis set consistency and it affects, particularly, to the short-range description.  A systematic improvement of the basis set does not guarantee a better wavefunction (see, for instance, the ground state geometries for different methods and basis sets in the NIST database).~\cite{NIST} In this study, we have optimized an even-tempered basis set of $n$s, $n$p and $n$d functions to perform the calculations of Be-like ions, where $n$ represents the number of orbitals used of each kind (these basis sets are referred as $n$SPD hereafter). \cite{matito:10pccp} Figure~\ref{fig:bz_convergence} displays the Coulomb hole $h_{\text{c}} (s)$ of the beryllium atom calculated with 6SPD, 7SPD, 8SPD, and 10SPD basis sets (9SPD is not shown due to its high resemblance to 10SPD). $h_{\text{c}} (s)$ with Dunning's augmented basis functions, aug-cc-pVDZ and aug-cc-pVTZ, and Pople's 6-311G** have also been calculated for comparison. Note that, as the quality of the basis set increases (from aug-cc-pVDZ to aug-cc-pVTZ), the short range part of $h_{\text{c}} (s)$ shows a sinusoidal shape that all  even-tempered basis sets already reproduce (see the bottom plot in Figure~\ref{fig:bz_convergence}). In a previous study, our group found that 6-311G** offers a poor description of the short-range part of the Coulomb hole, and that optimized basis sets provided a better description for core electrons.~\cite{rodriguez:19co} Therefore, we have chosen the 10SPD basis set to study the rest of Be-like ions as the 9SPD one was indistinguishable from 10SPD (\textit{i.e.} the basis set is converged with respect to the number of $s$, $p$ and $d$ basis functions).\newline

\section{The short-range region of the Coulomb hole of Be$_2$}

\begin{figure}[h!]
  \centering
  \includegraphics[width=0.80\linewidth]{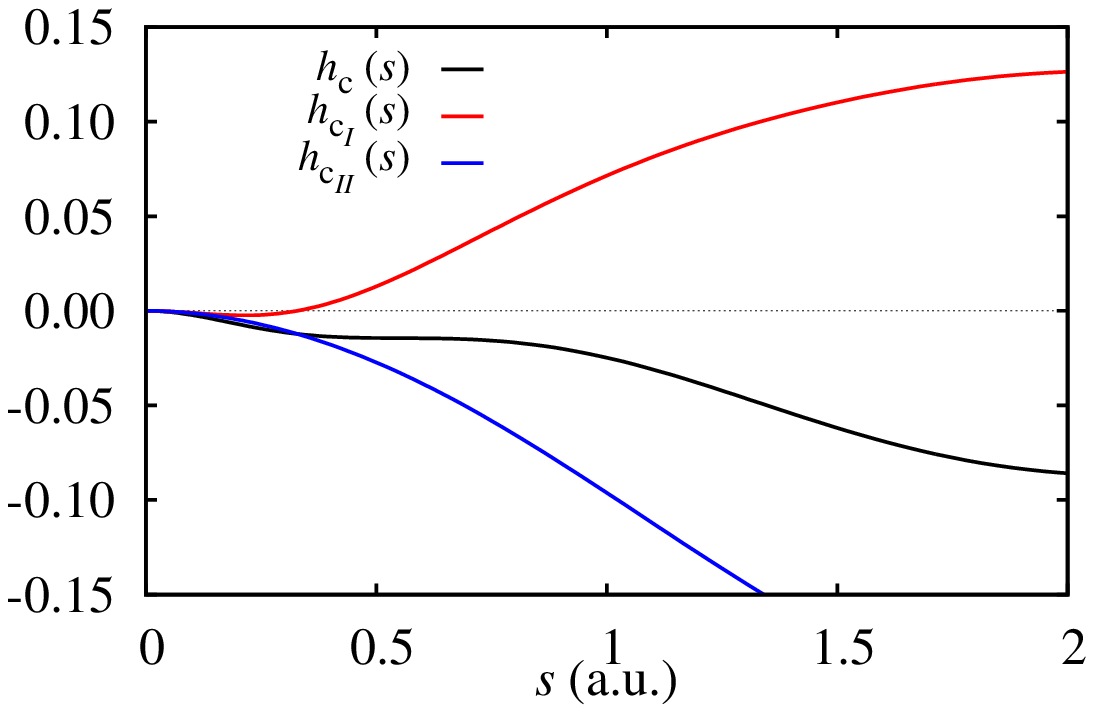} 
  \caption{The short-range region of the Coulomb hole ($h_{\text{c}} (s)$), $h_{\text{c}_{\textit{I}}} (s)$, and $h_{\text{c}_{\textit{II}}} (s)$ of Be$_2$ at 24.57 a.u. of bond distance.}
  \label{fig:be2_colze}
\end{figure}

The short-range region of the Coulomb hole of the beryllium dimer does not present a smooth decrease as the rest of Coulomb holes exposed in the article. Figure \ref{fig:be2_colze} contains this short-range region for a stretched geometry of Be$_2$, where the Coulomb hole $h_{\text{c}} (s)$ has an inflection point instead of a smooth minimum, which is usually defined as a shoulder-shape profile. The correlation decomposition of the Coulomb hole indicates that $h_{\text{c}_I} (s)$ is zero in the short-range region, with a posterior even growth, and $h_{\text{c}_{\textit{II}}} (s)$ decreases smoothly with the interelectronic distance $s$. The combination of both components results in a shoulder-like shape in the short-range region of $h_{\text{c}}$. 

\section{Type A nondynamic correlation Coulomb hole of stretched H$_2$}

\begin{figure}[h!]
  \centering
  \includegraphics[width=0.8\linewidth]{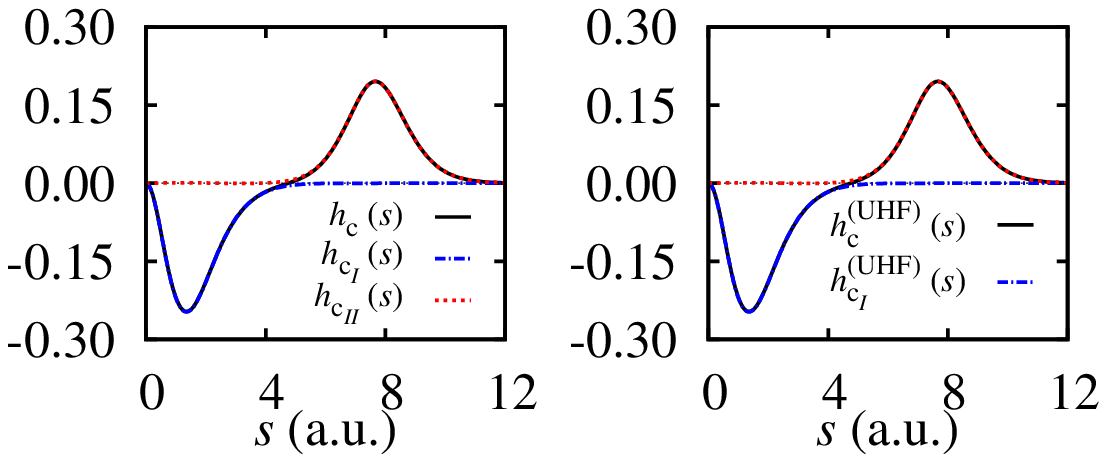} 
  \caption{(Left) Coulomb hole $h_{\text{c}} (s)$, $h_{\text{c}_{\textit{I}}} (s)$, and $h_{\text{c}_{\textit{II}}} (s)$ of H$_2$ at $R = 7.56$ a.u. in a minimal basis set, as presented in Figure~2 of the manuscript. (Right) Coulomb hole calculated using the UHF intracule pair density instead of the FCI one: $h_{\text{c}}^{\text{UHF}} (s) = I(\rho_2^{\text{UHF}},s) - I(\rho_2^{\text{RHF}},s)$, and $h_{\text{c}_{\textit{II}}}^{\text{UHF}} (s) = I(\rho_2^{\text{UHF}},s) - I(\rho_2^{\text{SD}},s)$.}
  \label{fig:uhf_typeA}
\end{figure}

Because UHF has the ability to describe type A nondynamic correlation, molecular dissociations such as the one of H$_2$ are correctly accounted for by this method. There is no type A nondynamic correlation in the equilibrium geometry of H$_2$, where we have not yet reached the Coulson-Fisher point. On the other hand, the bond stretching gives rise to type A nondynamic correlation and, eventually, the UHF intracule in a minimal basis set is exactly coincident with the FCI one. Thus, the Coulomb holes are identical, as depicted in Figure~\ref{fig:uhf_typeA}.

\section{Coulomb hole of He-Ne Atomic series}

\begin{figure}[h!]
  \centering
  \includegraphics[width=0.8\linewidth]{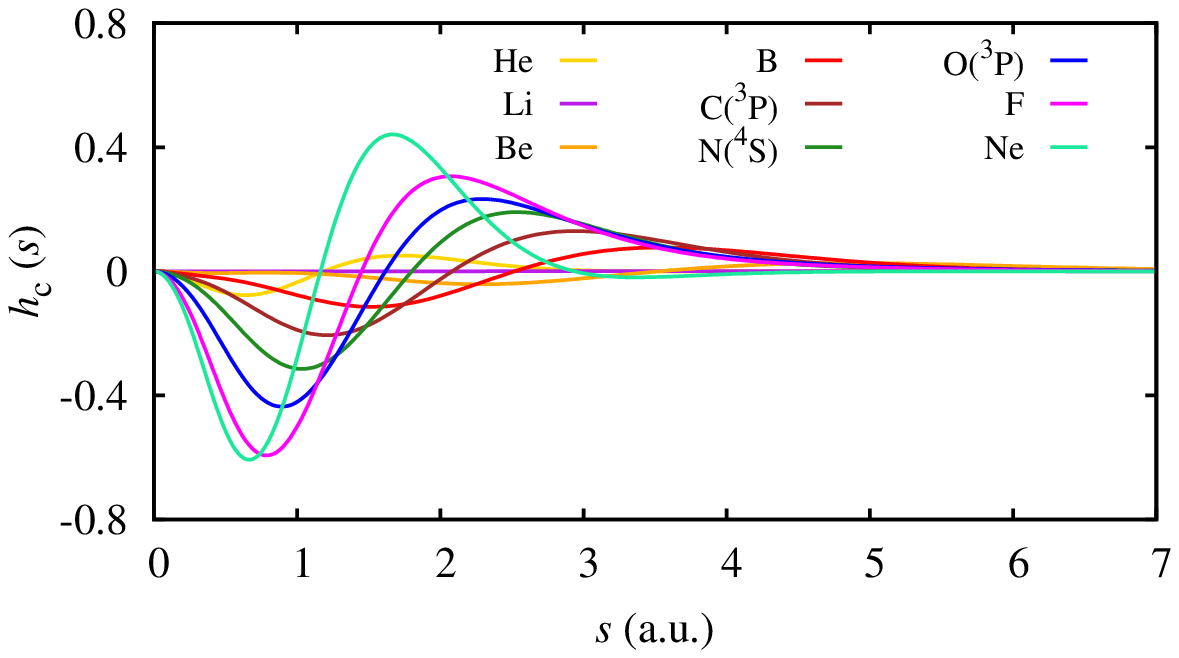} 
  \caption{Coulomb hole $h_{\text{c}} (s)$ of the He-Ne atomic series.}
  \label{fig:atoms}
\end{figure}

Coulomb holes $h_{\text{c}} (s)$ for the He-Ne atomic series in their ground state are presented in Figure~\ref{fig:atoms}. Notice that the dimension of lithium's Coulomb hole is smaller compared to the rest of atomic series and shows as a flat line.\newline

\section{Cases with RHF dissociating to an incorrect number of electrons per fragment.}

\begin{figure}[h!]
  \centering
  \includegraphics[width=0.65\linewidth]{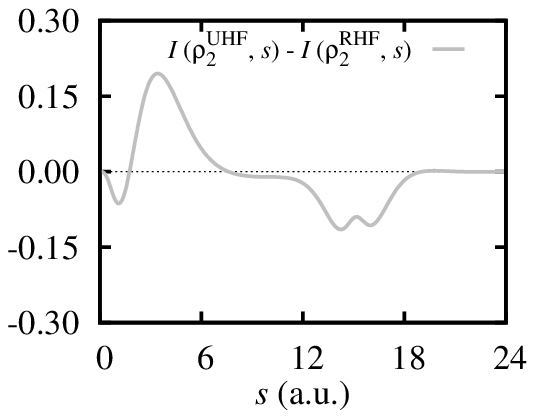} 
  \caption{The intracule probability density difference of the UHF and RHF pair densities, $I(\rho_2^{\text{UHF}},s) - I(\rho_2^{\text{RHF}},s)$, of the LiH molecule at $R = 15.12$ a.u..}
  \label{fig:lih_uhfvsrhf}
\end{figure}

Because RHF is not able to properly dissociate heterogeneous diatomic molecules with different electronegativities, the dissociated fragments of LiH are not neutral; namely, RHF dissociates the molecule into ions, Li+ and H-. Because of this, the RHF intracule probability density unusually presents an excess of electron-pair probability at the long-ranges of interelectronic distance caused by the extra electron in H. Instead, the unrestricted formalism correctly separates LiH into two neutral atoms. The UHF wavefunction (and also the intracule probability density) provides a close description to the FCI one. Figure \ref{fig:lih_uhfvsrhf} contains the intracule difference between the UHF and RHF. This difference is indistinguishable from the Coulomb hole $h_{\text{c}}$ presented in the manuscript, due to the quality of the UHF description. \newline

A similar situation occurs in the trigonal planar H$_4$ molecule (C$_{3v}$ symmetry point group). The HF description describes the axial hydrogen with a non-integer number of electrons. The UHF wavefunction provides a better description of the molecule, and hence the magnitude of the corresponding hole is smaller. 

\begin{figure}[h!]
  \centering
  \includegraphics[width=0.65\linewidth]{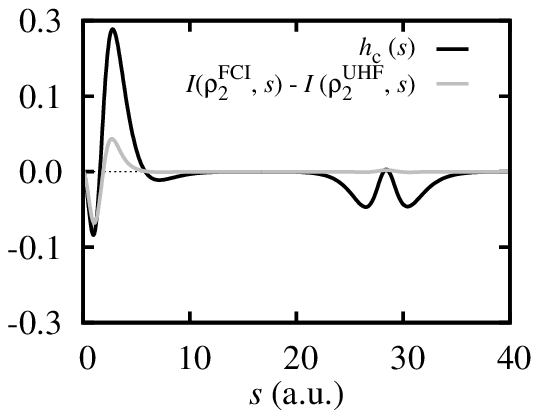} 
  \caption{Coulomb hole $h_{\text{c}}$ (black) and the intracule difference between the FCI and the UHF pair densities $I(\rho_2^{\text{FCI}},s) - I(\rho_2^{\text{UHF}},s)$ (gray) of H$_4$ molecule at C$_{3v}$ symmetry point group, with $_{\text{H}_{\Delta}} = 1.77$ a.u., and H$_{\text{ax}}$ placed at $R = 28.35$ a.u. from the equatorial plane (see also Figure~11 of the paper).}
  \label{fig:h4_uhfvsrhf}
\end{figure}

\section{Asymptotics of $\Delta\rho_2^{\text{c}_{\textit{I}}} (\boldsymbol{1},\boldsymbol{2})$ and $\Delta\rho_2^{\text{c}_{\textit{II}}} (\boldsymbol{1},\boldsymbol{2})$: H$_2$ in a minimal basis}

Let us analyze $\Delta\rho_2^{\text{c}_{\textit{I}}} (\boldsymbol{1},\boldsymbol{2})$ and $\Delta\rho_2^{\text{c}_{\textit{II}}} (\boldsymbol{1},\boldsymbol{2})$ for the hydrogen dimer described in a minimal basis set with a normalized 1$s$ orbital centered on nuclei $A$ and $B$. The bonding and antibonding restricted orbitals that arise are determined by the symmetry,
\begin{equation}\label{eqn:phi1}
\phi_1\left(\boldsymbol{r}\right)=\frac{1}{\sqrt{2\left(1+{s_{AB}}\right)}} \left(s_A (\boldsymbol{r}) + s_B (\boldsymbol{r}) \right),
\end{equation}
and
\begin{equation}
\phi_2\left(\boldsymbol{r}\right)=\frac{1}{\sqrt{2\left(1-{s_{AB}}\right)}} \left(s_A (\boldsymbol{r}) - s_B (\boldsymbol{r}) \right),
\end{equation}
respectively, and are eigenfunctions of the Fock operator. $s_{AB}$ is the overlap between both atomic orbitals.\newline

The Slater determinant built with the lowest-lying molecular orbital, eq~\ref{eqn:phi1}, corresponds to the restricted Hartree-Fock ground-state, $\Phi_0 \left(\boldsymbol{1},\boldsymbol{2}\right)$:
\begin{equation}
\Phi_0 \left(\boldsymbol{1},\boldsymbol{2}\right) = \vert1\overline{1} \rangle = \frac{1}{\sqrt{2}} \left( \phi_1\left(\boldsymbol{1}\right) \overline{\phi_1}\left(\boldsymbol{2}\right) + \overline{\phi_1}\left(\boldsymbol{1}\right)\phi_1\left(\boldsymbol{2}\right) \right),
\end{equation}
where the bar over the atomic orbital indicates that the spin state of the electron is beta (otherwise, it is alpha). The doubly-excited Slater determinant $\Phi_{\text{D}} \left(\boldsymbol{1},\boldsymbol{2}\right)$ reads
\begin{equation}
\Phi_{\text{D}} \left(\boldsymbol{1},\boldsymbol{2}\right) = \vert2\overline{2} \rangle = \frac{1}{\sqrt{2}} \left( \phi_2\left(\boldsymbol{1}\right) \overline{\phi_2}\left(\boldsymbol{2}\right) + \overline{\phi_2}\left(\boldsymbol{1}\right)\phi_2\left(\boldsymbol{2}\right) \right),
\end{equation}
and with a linear combination of both determinants one can construct the exact wavefunction for H$_2$ in a minimal basis set:
\begin{equation}
\Psi \left(\boldsymbol{1},\boldsymbol{2}\right) = c_0 \Phi_0 \left(\boldsymbol{1},\boldsymbol{2}\right) + c_{\text{D}} \Phi_{\text{D}} \left(\boldsymbol{1},\boldsymbol{2}\right),
\end{equation}
and, to keep the wavefunction normalized we impose $c_0^2 + c_{\text{D}}^2 = 1$.\newline

From this wavefunction, one can obtain its second-order reduced density matrix (2-RDM),
\begin{equation}
\begin{split}
\rho_2\left(\boldsymbol{1},\boldsymbol{2};\boldsymbol{1'},\boldsymbol{2'}\right) = 
2 \Psi^* & \left(\boldsymbol{1},\boldsymbol{2}\right) \Psi \left(\boldsymbol{1'},\boldsymbol{2'}\right) \\
= c_0^2 \Bigl(&
\phi_1^*(\boldsymbol{1})
\overline{\phi_1^*}(\boldsymbol{2})
\phi_1(\boldsymbol{1'})
\overline{\phi_1}(\boldsymbol{2'})
+ 
\overline{\phi_1^*}(\boldsymbol{1})
\phi_1^*(\boldsymbol{2})
\overline{\phi_1}(\boldsymbol{1'})
\phi_1(\boldsymbol{2'}) \\
&+
\phi_1^*(\boldsymbol{1})
\overline{\phi_1^*}(\boldsymbol{2})
\overline{\phi_1}(\boldsymbol{1'})
\phi_1(\boldsymbol{2'})
+ 
\overline{\phi_1^*}(\boldsymbol{1})
\phi_1^*(\boldsymbol{2})
\phi_1(\boldsymbol{1'})
\overline{\phi_1}(\boldsymbol{2'}) 
\Bigr)\\
+c_{\text{D}}^2 \Bigl(&
\phi_2^*(\boldsymbol{1})
\overline{\phi_2^*}(\boldsymbol{2})
\phi_2(\boldsymbol{1'})
\overline{\phi_2}(\boldsymbol{2'})
+ 
\overline{\phi_2^*}(\boldsymbol{1})
\phi_2^*(\boldsymbol{2})
\overline{\phi_2}(\boldsymbol{1'})
\phi_2(\boldsymbol{2'}) \\
&+
\phi_2^*(\boldsymbol{1})
\overline{\phi_2^*}(\boldsymbol{2})
\overline{\phi_2}(\boldsymbol{1'})
\phi_2(\boldsymbol{2'}) 
+ 
\overline{\phi_2^*}(\boldsymbol{1})
\phi_2^*(\boldsymbol{2})
\phi_2(\boldsymbol{1'})
\overline{\phi_2}(\boldsymbol{2'}) 
\Bigr)\\
+c_0 c_{\text{D}} \Bigl(&
\phi_1^*(\boldsymbol{1})
\overline{\phi_1^*}(\boldsymbol{2})
\phi_2(\boldsymbol{1'})
\overline{\phi_2}(\boldsymbol{2'}) 
+
\overline{\phi_1^*}(\boldsymbol{1})
\phi_1^*(\boldsymbol{2})
\overline{\phi_2}(\boldsymbol{1'})
\phi_2(\boldsymbol{2'})\\
&+
\phi_1^*(\boldsymbol{1})
\overline{\phi_1^*}(\boldsymbol{2})
\overline{\phi_2}(\boldsymbol{1'})
\phi_2(\boldsymbol{2'})
+
\overline{\phi_1^*}(\boldsymbol{1})
\phi_1^*(\boldsymbol{2})
\phi_2(\boldsymbol{1'})
\overline{\phi_2}(\boldsymbol{2'})
\\
&+
\phi_2^*(\boldsymbol{1})
\overline{\phi_2^*}(\boldsymbol{2})
\phi_1(\boldsymbol{1'})
\overline{\phi_1}(\boldsymbol{2'})
+
\overline{\phi_2^*}(\boldsymbol{1})
\phi_2^*(\boldsymbol{2})
\overline{\phi_1}(\boldsymbol{1'})
\phi_1(\boldsymbol{2'}) \\
&+
\phi_2^*(\boldsymbol{1})
\overline{\phi_2^*}(\boldsymbol{2})
\overline{\phi_1}(\boldsymbol{1'})
\phi_1(\boldsymbol{2'})
+
\overline{\phi_2^*}(\boldsymbol{1})
\phi_2^*(\boldsymbol{2})
\phi_1(\boldsymbol{1'})
\overline{\phi_1}(\boldsymbol{2'})
\Bigr);
\end{split}
\label{eqn:2rdm}
\end{equation}
the first-order reduced density matrix (1-RDM), which is obtained upon integration of the 2-RDM over the second-electron coordinate,
\begin{equation}
\begin{split}
\rho_1\left(\boldsymbol{1};\boldsymbol{1'}\right) &= \int{d\boldsymbol{2}\hspace{0.1cm}
\rho_2\left(\boldsymbol{1},\boldsymbol{2};\boldsymbol{1'},\boldsymbol{2'}\right)
\vert_{\boldsymbol{2'}=\boldsymbol{2}}} \\
&= 
c_0^2 \left(
\phi_1^*(\boldsymbol{1})\phi_1(\boldsymbol{1'})
+
\overline{\phi_1^*}(\boldsymbol{1})\overline{\phi_1}(\boldsymbol{1'})
\right)
+ c_{\text{D}}^2 \left(
\phi_2^*(\boldsymbol{1})\phi_2(\boldsymbol{1'})
+
\overline{\phi_2^*}(\boldsymbol{1})\overline{\phi_2}(\boldsymbol{1'})
\right);
\end{split}
\end{equation}
and the electron density, which is the diagonal of the 1-RDM, 
\begin{equation}
\begin{split}
\rho(\boldsymbol{1}) = \rho_1\left(\boldsymbol{1};\boldsymbol{1}\right) 
=
c_0^2 \left( \vert\phi_1(\boldsymbol{1})\vert^2 + \vert\overline{\phi_1}(\boldsymbol{1})\vert^2 \right)+
c_{\text{D}}^2 \left( \vert\phi_2(\boldsymbol{1})\vert^2 + \vert\overline{\phi_2}(\boldsymbol{1})\vert^2 \right).
\end{split}
\end{equation}

The HF pair density is
\begin{equation}
\rho_2^{\text{HF}}(\boldsymbol{1},\boldsymbol{2}) = 
2 \Phi_0^*\left(\boldsymbol{1},\boldsymbol{2}\right) \Phi_0 \left(\boldsymbol{1},\boldsymbol{2}\right) =
\vert\phi_1(\boldsymbol{1})\vert^2 \vert\overline{\phi_1}(\boldsymbol{2})\vert^2+
\vert\overline{\phi_1}(\boldsymbol{1})\vert^2 \vert\phi_1(\boldsymbol{2})\vert^2
\end{equation}
and the single-determinant approximation, according to eq~6 in the manuscript,~\cite{lowdin:55pr} reads
\begin{equation}
\begin{split}
\rho_2^{\text{SD}}(\boldsymbol{1},\boldsymbol{2}) =& 
\rho(\boldsymbol{1})\rho(\boldsymbol{2}) - \vert\rho_1(\boldsymbol{1};\boldsymbol{2})\vert^2\\
=&c_0^4 \left( 
\vert\phi_1(\boldsymbol{1})\vert^2 \vert\overline{\phi_1}(\boldsymbol{2})\vert^2+
\vert\overline{\phi_1}(\boldsymbol{1})\vert^2\vert\phi_1(\boldsymbol{2})\vert^2\right)+
c_{\text{D}}^4 \left(
\vert\phi_2(\boldsymbol{1})\vert^2 \vert\overline{\phi_2}(\boldsymbol{2})\vert^2+
\vert\overline{\phi_2}(\boldsymbol{1})\vert^2\vert\phi_2(\boldsymbol{2})\vert^2\right)\\
+&c_0^2c_{\text{D}}^2 \biggl(
\vert\phi_1(\boldsymbol{1})\vert^2\vert\overline{\phi_2}(\boldsymbol{2})\vert^2+
\vert\overline{\phi_1}(\boldsymbol{1})\vert^2\vert\phi_2(\boldsymbol{2})\vert^2+
\vert{\phi_1}(\boldsymbol{1})\vert^2\vert\phi_2(\boldsymbol{2})\vert^2+
\vert\overline{\phi_1}(\boldsymbol{1})\vert^2\vert\overline{\phi_2}(\boldsymbol{2})\vert^2\\
+&\vert\phi_1(\boldsymbol{2})\vert^2\vert\overline{\phi_2}(\boldsymbol{1})\vert^2+
\vert\overline{\phi_1}(\boldsymbol{2})\vert^2\vert\phi_2(\boldsymbol{1})\vert^2+
\vert{\phi_1}(\boldsymbol{2})\vert^2\vert\phi_2(\boldsymbol{1})\vert^2+
\vert\overline{\phi_1}(\boldsymbol{2})\vert^2\vert\overline{\phi_2}(\boldsymbol{1})\vert^2\\
-&\phi_1^{*}(\boldsymbol{1})\phi_1(\boldsymbol{2})\phi_2^{*}(\boldsymbol{1})\phi_2(\boldsymbol{2})-
\overline{\phi_1^{*}}(\boldsymbol{1})\overline{\phi_1}(\boldsymbol{2})\overline{\phi_2^{*}}(\boldsymbol{1})\overline{\phi_2}(\boldsymbol{2})\\
-&\phi_1^{*}(\boldsymbol{2})\phi_1(\boldsymbol{1})\phi_2^{*}(\boldsymbol{2})\phi_2(\boldsymbol{1})-
\overline{\phi_1^{*}}(\boldsymbol{2})\overline{\phi_1}(\boldsymbol{1})\overline{\phi_2^{*}}(\boldsymbol{2})\overline{\phi_2}(\boldsymbol{1}) \biggr).
\end{split}
\label{eqn:SDpd}
\end{equation}

At the particular case when the two centers are far apart (and electrons are located at each center), the overlap between the atomic orbitals becomes zero, \textit{i.e.}, $\vert R_{AB} \vert \to \infty \Rightarrow s_{AB} \to 0$. Then, it is legitimate to consider that $\phi_1(A)\approx\phi_2(A)$ and $-\phi_1(B)\approx \phi_2(B)$, $A$ and $B$ standing for the position of the corresponding atom. Therefore, eq~\ref{eqn:SDpd} becomes
\begin{equation}\label{fig:s11}
\begin{split}
\lim_{\vert R_{AB}\vert\to\infty}\rho_2^{\text{SD}}\left(A,B\right) &= 
\left(c_0^4 + c_{\text{D}}^4 + 2c_0^2c_{\text{D}}^2\right) 
\left(\vert\phi_1(A)\vert^2\vert\overline{\phi_1}(B)\vert^2 
+\vert\overline{\phi_1}(A)\vert^2\vert\phi_1(B)\vert^2\right)\\
&+2c_0^2c_{\text{D}}^2
\biggl(\vert\phi_1(A)\vert^2\vert\phi_1(B)\vert^2+\vert\overline{\phi_1}(A)\vert^2\vert\overline{\phi_1}(B)\vert^2\\
&+\phi_1^{*}(A)\phi_1(B)\phi_1^{*}(A)\phi_1(B)
+\overline{\phi_1^{*}}(A)\overline{\phi_1}(B)\overline{\phi_1^{*}}(A)\overline{\phi_1}(B)
\biggr)\\
&=\rho_2^{\text{HF}}\left(A,B\right) + 4c_0^2c_{\text{D}}^2
\biggl(\vert\phi_1(A)\vert^2\vert\phi_1(B)\vert^2+\vert\overline{\phi_1}(A)\vert^2\vert\overline{\phi_1}(B)\vert^2\biggr),
\end{split}
\end{equation}
where, for the sake of simplicity, we have considered real orbitals, $\phi_i^{*}(\boldsymbol{r}) = \phi_i(\boldsymbol{r})$. Hence,
\begin{equation}
\begin{split}
\lim_{\vert R_{AB} \vert \to \infty}\Delta\rho_2^{\text{c}_{\textit{I}}}(A,B) = 
\lim_{\vert R_{AB} \vert \to \infty} \left[\rho_2^{\text{SD}}(A,B)-\rho_2^{\text{HF}}(A,B) \right]
= 4c_0^2c_{\text{D}}^2
\biggl(\vert\phi_1(A)\vert^2\vert\phi_1(B)\vert^2+\vert\overline{\phi_1}(A)\vert^2\vert\overline{\phi_1}(B)\vert^2\biggr).
\end{split}
\end{equation}

When the wavefunction is only composed of the HF determinant, $c_{\text{D}} = 0 \Rightarrow \Delta\rho_2^{\text{c}_{\textit{I}}}(A,B) \to 0$. Conversely, at the limit of the dissociation, the expansion coefficients are equal in weight, $c_0 = - c_{\text{D}} = \frac{1}{\sqrt{2}}$. Hence, the spin-integrated $\text{c}_{\textit{I}}$ component of the pair density tends to a value which is as large as the HF pair density, $\Delta\rho_2^{\text{c}_{\textit{I}}}(A,B) \to \rho_2^{\text{HF}}(A,B)$, which results in $\Delta\rho_2^{\text{c}_{\textit{I}}}(A,B)$ capturing the nondynamic correlation arisen from the stretching. Hence, $\Delta\rho_2^{\text{c}_{\textit{I}}}$ corrects the RHF pair density by adding twice as many electron pairs as RHF at large distances.
\newline

One could also analyze the spin-components of these expressions. The hydrogen molecule only presents opposite-spin interactions, with just one alpha and one beta electron. Thus the HF pair density does not have exchange terms and will read
\begin{equation}
\rho_2^{\text{HF}}(\boldsymbol{1},\boldsymbol{2}) = \rho^{{\text{HF}},\alpha}(\boldsymbol{1})\rho^{{\text{HF}},\,\beta}(\boldsymbol{2})+\rho^{{\text{HF}},\,\beta}(\boldsymbol{1})\rho^{{\text{HF}},\alpha}(\boldsymbol{2}).
\end{equation}
However, $\rho_2^{\text{SD}}$ includes some same-spin elements by construction, and hence
\begin{equation}
\rho_2^{\text{SD}} (\boldsymbol{1},\boldsymbol{2})=
\rho^{\alpha}(\boldsymbol{1})\rho^{\alpha}(\boldsymbol{2})+
\rho^{\beta}(\boldsymbol{1})\rho^{\beta}(\boldsymbol{2})+
\rho^{\alpha}(\boldsymbol{1})\rho^{\beta}(\boldsymbol{2})+
\rho^{\beta}(\boldsymbol{1})\rho^{\alpha}(\boldsymbol{2})-
\vert \rho_1^{\alpha\alpha}(\boldsymbol{1};\boldsymbol{2})\vert^2-
\vert \rho_1^{\beta\beta}(\boldsymbol{1};\boldsymbol{2})\vert^2.
\end{equation}
If we assume $\rho^{\text{HF},\sigma}(\boldsymbol{1})\approx\rho^{\sigma}(\boldsymbol{1})$, only the same-spin terms survive in the c$_{\textit{I}}$ component,
\begin{equation}
\Delta\rho_2^{\text{c}_{\textit{I}}}(\boldsymbol{1},\boldsymbol{2}) \approx
\rho^{\alpha}(\boldsymbol{1})\rho^{\alpha}(\boldsymbol{2})+
\rho^{\beta}(\boldsymbol{1})\rho^{\beta}(\boldsymbol{2})-
\vert \rho_1^{\alpha\alpha}(\boldsymbol{1};\boldsymbol{2})\vert^2-
\vert \rho_1^{\beta\beta}(\boldsymbol{1};\boldsymbol{2})\vert^2.
\end{equation}
When the dissociation occurs, namely when $R_{AB} \to \infty$,  $\Delta\rho_2^{\text{c}_{\textit{I}}}(\boldsymbol{1},\boldsymbol{2})$ vanishes at short range (the first non-vanishing term of the short-range expansion is quadratic in $r_{12}$).\newline

The asymptotics of the 1-RDM define the asymptotics of the $\text{c}_I$ component of the pair density. March and Pucci~\cite{march:81jcp} found that, when electrons are separated infinitely from each other and are also separated from any nucleus, $\rho_1(\boldsymbol{1};\boldsymbol{2})\to\sqrt{\rho(\boldsymbol{1})\rho(\boldsymbol{2})}$.\footnote{For non-degenerate $(N-1)$-particle systems.~\cite{ernzerhof:96jcp}} Therefore, the same-spin component of the SD approximation reduces to zero, and thus 
\begin{equation}
\lim_{{r_{12} \to \infty}\atop{r_{A1},r_{B2}\to \infty}} \Delta\rho_2^{\text{c}_{\textit{I}},\alpha\alpha} (\boldsymbol{1},\boldsymbol{2})= 0
\hspace{3cm}\forall A,B
\end{equation}
The opposite-spin component does not depend on the 1-RDM, therefore its asymptotics is defined by the difference between the product of the correlated and uncorrelated electron densities:
\begin{equation}
\lim_{{r_{12} \to \infty}\atop{r_{A1},r_{B2}\to \infty}} \Delta\rho_2^{\text{c}_{\textit{I}},\alpha\beta} (\boldsymbol{1},\boldsymbol{2})= \rho^{\alpha}(\boldsymbol{1})\rho^{\beta}(\boldsymbol{2}) - \rho^{\text{HF},\alpha}(\boldsymbol{1})\rho^{\text{HF},\beta}(\boldsymbol{2})= 0,\hspace{2cm}\forall A,B
\end{equation}
because the density dies off quickly far from the nuclei. Some of the expressions we have developed thus far hold regardless of the size of the basis set, however, the latter equality is only strictly attained for a minimal basis set.\newline

When electrons are placed on top of each other ($r_{12}\to 0$, the coalescence point), the same-spin component of $\Delta\rho_2^{\text{c}_{\textit{I}}} (\boldsymbol{1},\boldsymbol{2})$ is zero by construction. Nevertheless, there is a small probability of opposite-spin electrons being on top of each other:
\begin{equation}
\lim_{r_{12}\to 0} \Delta\rho_2^{\text{c}_{\textit{I}},\alpha\beta} (\boldsymbol{1},\boldsymbol{2}) = \rho^{\alpha}(\boldsymbol{1})\rho^{\beta}(\boldsymbol{1}) - \rho^{\text{HF},\alpha}(\boldsymbol{1})\rho^{\text{HF},\beta}(\boldsymbol{1}).
\end{equation}

Now, let us consider the limits of $\Delta\rho_2^{\text{c}_{\textit{II}}} (\boldsymbol{1},\boldsymbol{2})$.
The value of $\Delta\rho_2^{\text{c}_{\textit{II}}} (\boldsymbol{1},\boldsymbol{2})$ at the coalescence point
depends exclusively on the opposite-spin component, because
the same-spin one vanishes due to the Pauli principle,
\begin{equation}
\lim_{r_{12}\to 0} \Delta\rho_2^{\text{c}_{\textit{II}}}(\boldsymbol{1},\boldsymbol{2})=\Delta\rho_2^{\text{c}_{\textit{II}},\alpha\beta} (\boldsymbol{1},\boldsymbol{1})+\Delta\rho_2^{\text{c}_{\textit{II}},\,\beta\alpha} (\boldsymbol{1},\boldsymbol{1}) = 2\left(\rho_2^{\alpha\beta}(\boldsymbol{1},\boldsymbol{1}) - \rho^{\alpha}(\boldsymbol{1})\rho^{\beta}(\boldsymbol{1})\right),
\end{equation}
and, hence, its behavior depends on the on-top pair density value.\cite{perdew:97ijqc}\newline

As $R_{AB}\to \infty$, the actual pair density (the diagonal elements of eq~\ref{eqn:2rdm}) is dominated by the value at the points close to the nuclei, 
\begin{equation}
\begin{split}
\rho_2(A,B) =& c_0^2 \left( \vert\phi_1(A)\vert^2\vert\overline{\phi_1}(B)\vert^2 
+
\vert\overline{\phi_1}(A)\vert^2\vert\phi_1(B)\vert^2
\right)
+ c_{\text{D}}^2 \left( \vert\phi_1(A)\vert^2\vert\overline{\phi_1}(B)\vert^2 
+
\vert\overline{\phi_1}(A)\vert^2\vert\phi_1(B)\vert^2
\right) \\
-& 2c_0c_{\text{D}} \left(
\vert\phi_1(A)\vert^2\vert\overline{\phi_1}(B)\vert^2 + 
\vert\overline{\phi_1}(A)\vert^2\vert\phi_1(B)\vert^2 
\right) \\
=&\left(c_0 - c_{\text{D}}\right)^2 \rho_2^{\text{HF}}(A,B),
\end{split}
\end{equation}
and therefore c$_\textit{II}$ component of the pair density is (see eq~\ref{fig:s11})
\begin{equation}
\Delta\rho_2^{\text{c}_{\textit{II}}}(A,B) = 
\rho_2(A,B)-\rho_2^{\text{SD}}(A,B) 
=
\left(c_0 - c_{\text{D}}\right)^2 \rho_2^{\text{HF}}(A,B) - (1 + 4c_0^2c_{\text{D}}^2)\rho_2^{\text{HF}}(A,B),
\end{equation}
where we have assumed a closed-shell system.

At the dissociation limit $c_0 = - c_{\text{D}} = \frac{1}{\sqrt{2}}$ and, therefore, $\lim_{R_{AB}\to \infty}\rho_2(A,B)= 2\rho_2^{\text{HF}}(A,B)$. Hence,  $\lim_{R_{AB}\to \infty}\Delta\rho_2^{\text{c}_{\textit{II}}} (A,B)= 0$. Unlike $\Delta\rho_2^{\text{c}_{\textit{I}}}(\boldsymbol{1},\boldsymbol{2})$, the long-range component of $\Delta\rho_2^{\text{c}_{\textit{II}}}(\boldsymbol{1},\boldsymbol{2})$ vanishes and, therefore, we have a convenient range separation of the pair density into $\text{c}_I$ and $\text{c}_{II}$ components.\newline

\bibliography{gen}